\documentclass[fleqn,10pt]{wlscirep}
\usepackage[utf8]{inputenc}
\usepackage[T1]{fontenc}
\usepackage{lineno}
\usepackage{multirow}
\usepackage{caption}
\usepackage{subcaption}
\usepackage{hyperref}

\title{The RETA Benchmark for Retinal Vascular Tree Analysis}

\author[1,*]{Xingzheng Lyu}
\author[2]{Li Cheng}
\author[1,*]{Sanyuan Zhang}
\affil[1]{College of Computer Science and Technology, Zhejiang University, Hangzhou, 310027, China.}
\affil[2]{Department of Electrical and Computer Engineering, University of Alberta, Edmonton, T6G 1H9, Canada.}

\affil[*]{corresponding author(s): Xingzheng Lyu (frankly@zju.edu.cn),Sanyuan Zhang (syzhang@zju.edu.cn)}

\begin{abstract}

Topological and geometrical analysis of retinal blood vessel is a cost-effective way for early detection of many common diseases. Meanwhile, automated vessel segmentation and vascular tree analysis are still lacking in terms of generalization capability. In this work, we construct a novel benchmark RETA with 81 labeled vessel masks aiming to facilitate retinal vessel analysis. A semi-automated coarse-to-fine workflow is proposed to annotating vessel pixels. During dataset construction, we strived to control inter-annotator variability and intra-annotator variability by performing multi-stage annotation and label disambiguation on self-developed dedicated software. In addition to binary vessel masks, we obtained vessel annotations containing artery/vein masks, vascular skeletons, bifurcations, trees and abnormalities during vessel labelling. Both subjective and objective quality validation of labeled vessel masks have demonstrated significant improved quality over other publicly datasets. The annotation software is also made publicly available for vessel annotation visualization. Users could develop vessel segmentation algorithms or evaluate vessel segmentation performance with our dataset. Moreover, our dataset might be a good research source for cross-modality tubular structure segmentation.

\end{abstract}
\begin{document}

\flushbottom
\maketitle

\thispagestyle{empty}


\section*{Background \& Summary}
Retinal vascular calibre is a biomarker to early detect microvascular and macrovascular diseases, such as diabetic retinopathy\cite{ikram2013retinal}, hypertensive retinopathy and stroke\cite{baker2008retinal}. Colour fundus image, a two-dimensional (2D) colour photograph captured by fundus camera, is a non-invasive way to evaluate retinal vascular abnormalities (e.g. calibre change, tortuosity alteration, neovascularization and arteriovenous nicking). Investigators attempted to use computer-aided methods that could automatically segment retinal blood vessels including arterioles and venules from retinal images \cite{cheng2019retinal}. Recent supervised deep learning methods classify every pixel within a fundus image into vessel/non-vessel and obtain state-of-the-art vessel segmentation performance on public datasets. Follow-up vessel analysis steps mainly aims to separate arteriolar/venular(A/V) trees from raw predictions. These steps include vessel skeletonization, vascular junctions (bifurcations and crossovers) identification, vascular tree tracking and vessel segment diameter measurement \cite{lyu2016construction}. Then, vascular calibre based quantitative metric, for example, arteriolar-to-venular diameter ratio\cite{knudtson2003revised}, is calculated to indicate vascular disease risk. However, cross-dataset evaluation shows robustness of segmentation model is still the bottleneck when predicting private images not in the training dataset\cite{laibacher2019evaluation}. Manual correction becomes inevitable before conducting above follow-up analysis.

Training set related factors that affect model robustness are the size of training set and the amount of noisy labels. To the best of our knowledge, there are more than ten public datasets for binary vessel segmentation. Some of them provides artery/vein labels. There are also datasets for vascular network analysis containing vascular keypoint detection \cite{abbasi2016automatic} and vascular tree tracking \cite{estrada2015retinal, zhao2019retinal}. Every independent tree starts from a starting vertex near optic disk (OD) and terminates at an ending vertex. Vascular junctions are internal vertices within a tree. Vascular network analysis is an effective technique for A/V tree separation and branching complexity analysis. These datasets are useful for developing automatic algorithms but far from enough. Image augmentation techniques, like image rotation and synthesis\cite{zhao2018synthesizing}, could increase the size of dataset. But for annotation noise, it is still unsolved because inter- and intra-annotator variabilities usually introduce noisy labels. This problem is frequently argued in public datasets\cite{trucco2013validating,yan2017skeletal}. Pixel-level vessel annotation is quite expensive and time-consuming. Only well-trained experts who use customed vessel annotation software, like VAMPIRE tool\cite{perez2011vampire}, equipped with user-friendly labelling tools could guarantee the quality of vessel annotations. 

\begin{figure}
     \centering
     \begin{subfigure}[b]{0.24\textwidth}
         \centering
         \includegraphics[width=\textwidth]{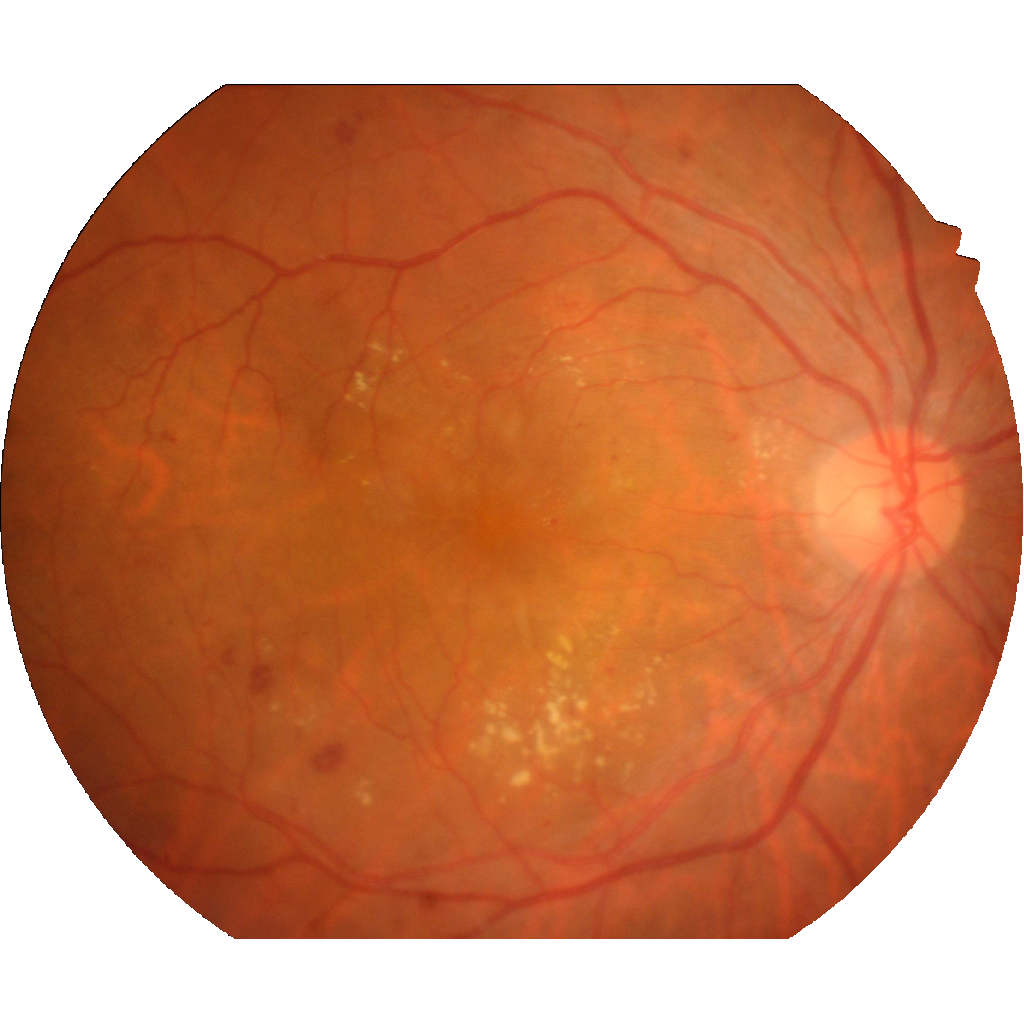}
         \caption{}
     \end{subfigure}
     \hfill
     \begin{subfigure}[b]{0.24\textwidth}
         \centering
         \includegraphics[width=\textwidth]{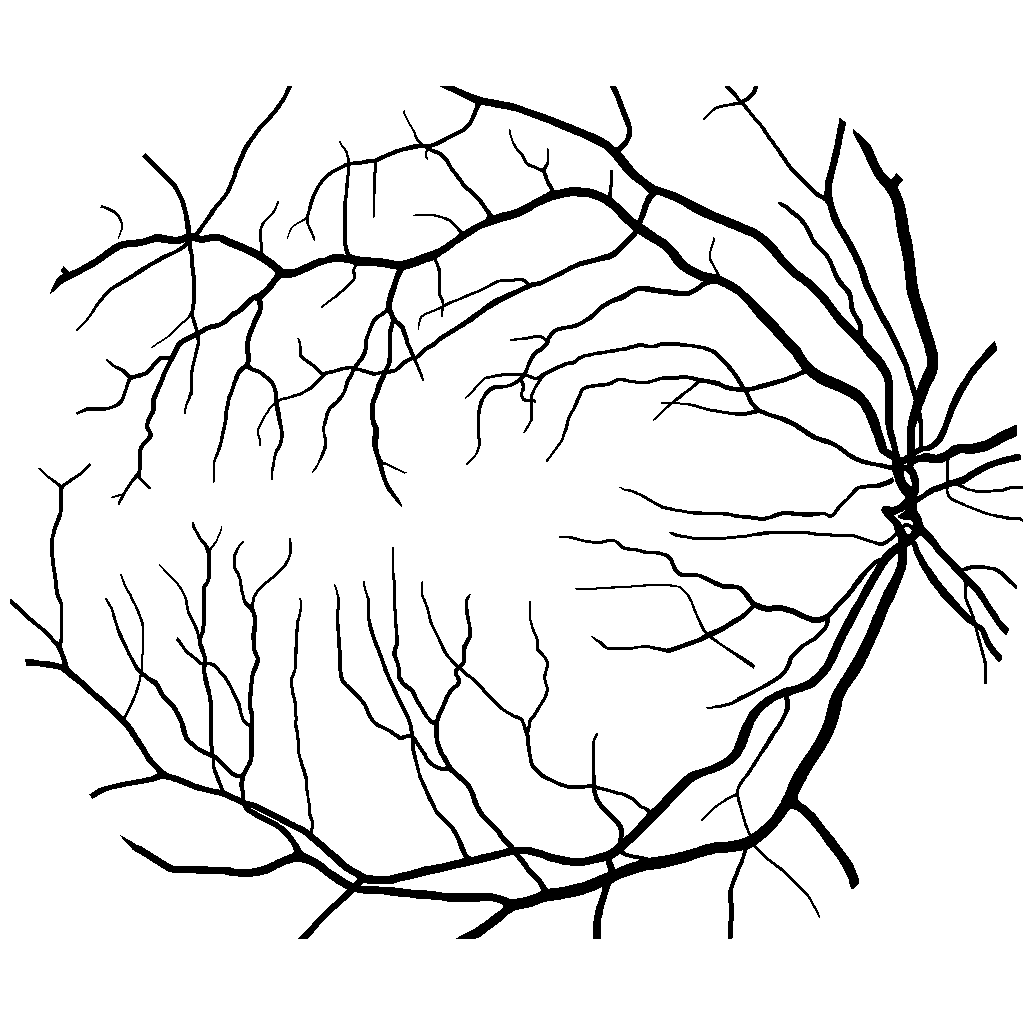}
         \caption{}
     \end{subfigure}
     \hfill
     \begin{subfigure}[b]{0.24\textwidth}
         \centering
         \includegraphics[width=\textwidth]{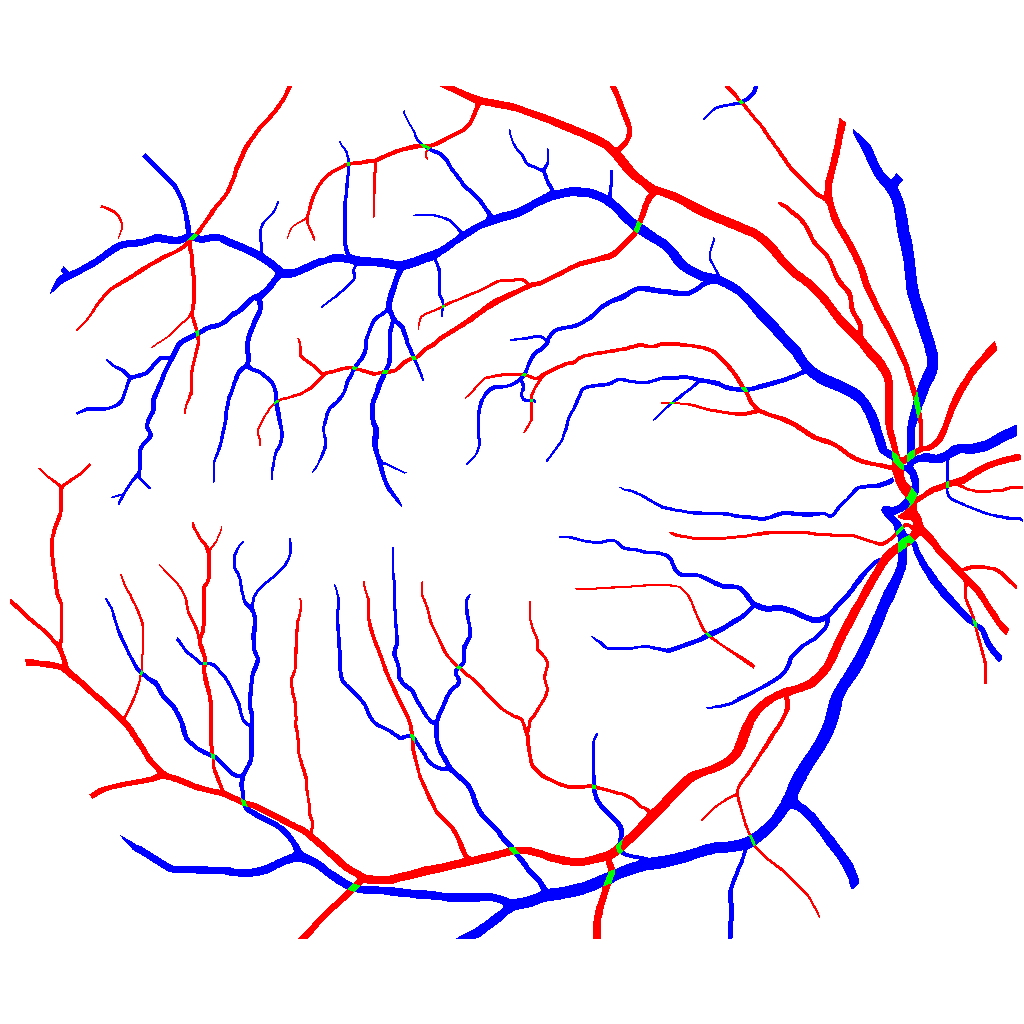}
         \caption{}
     \end{subfigure}
     \hfill
     \begin{subfigure}[b]{0.24\textwidth}
         \centering
         \includegraphics[width=\textwidth]{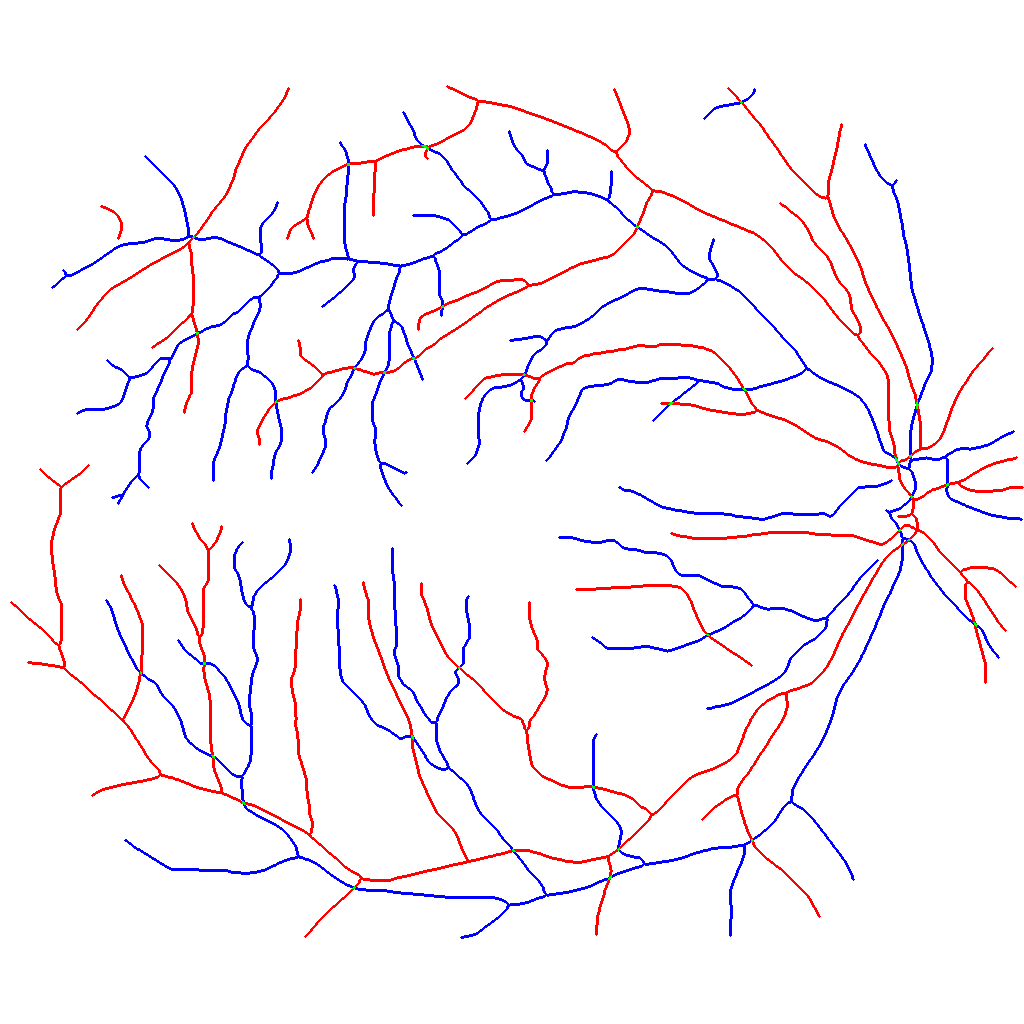}
         \caption{}
     \end{subfigure}
     \hfill
     \begin{subfigure}[b]{0.24\textwidth}
         \centering
         \includegraphics[width=\textwidth]{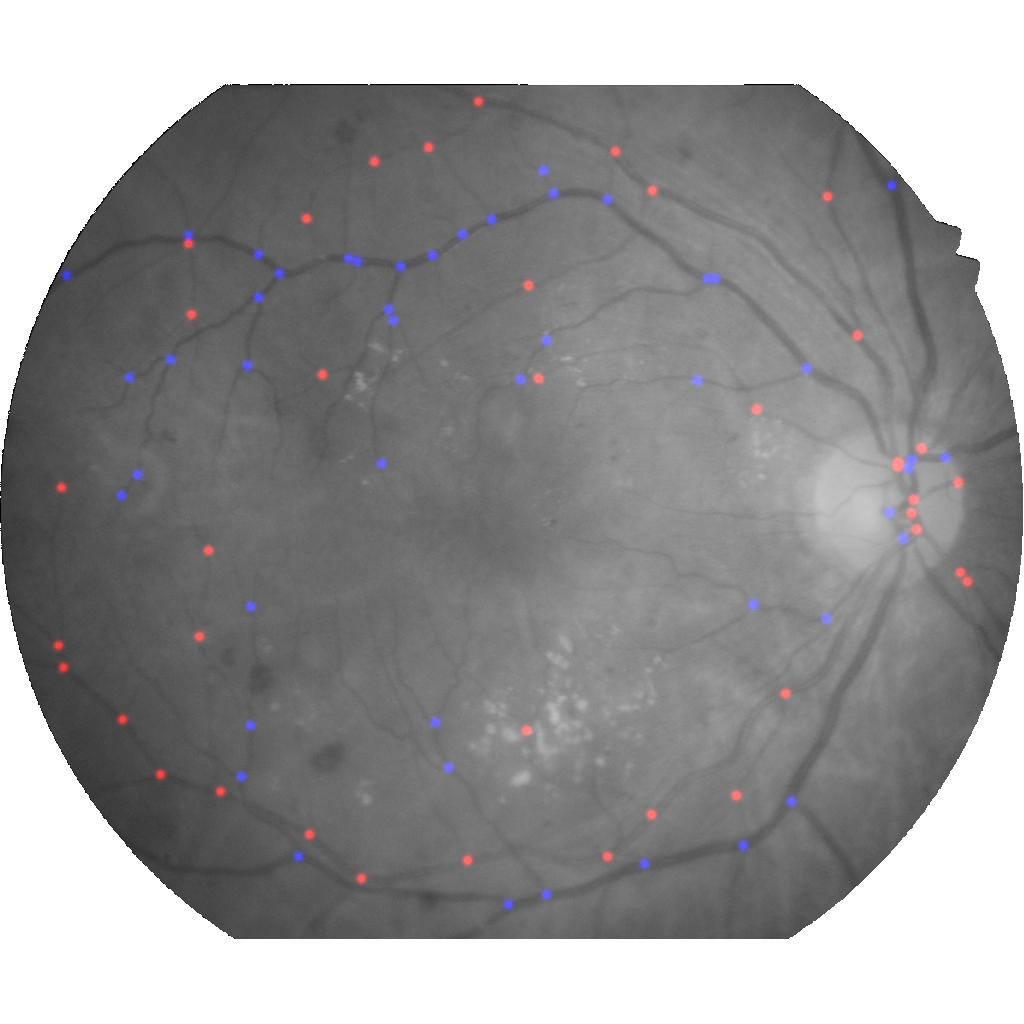}
         \caption{}
     \end{subfigure}
     \hfill
     \begin{subfigure}[b]{0.24\textwidth}
         \centering
         \includegraphics[width=\textwidth]{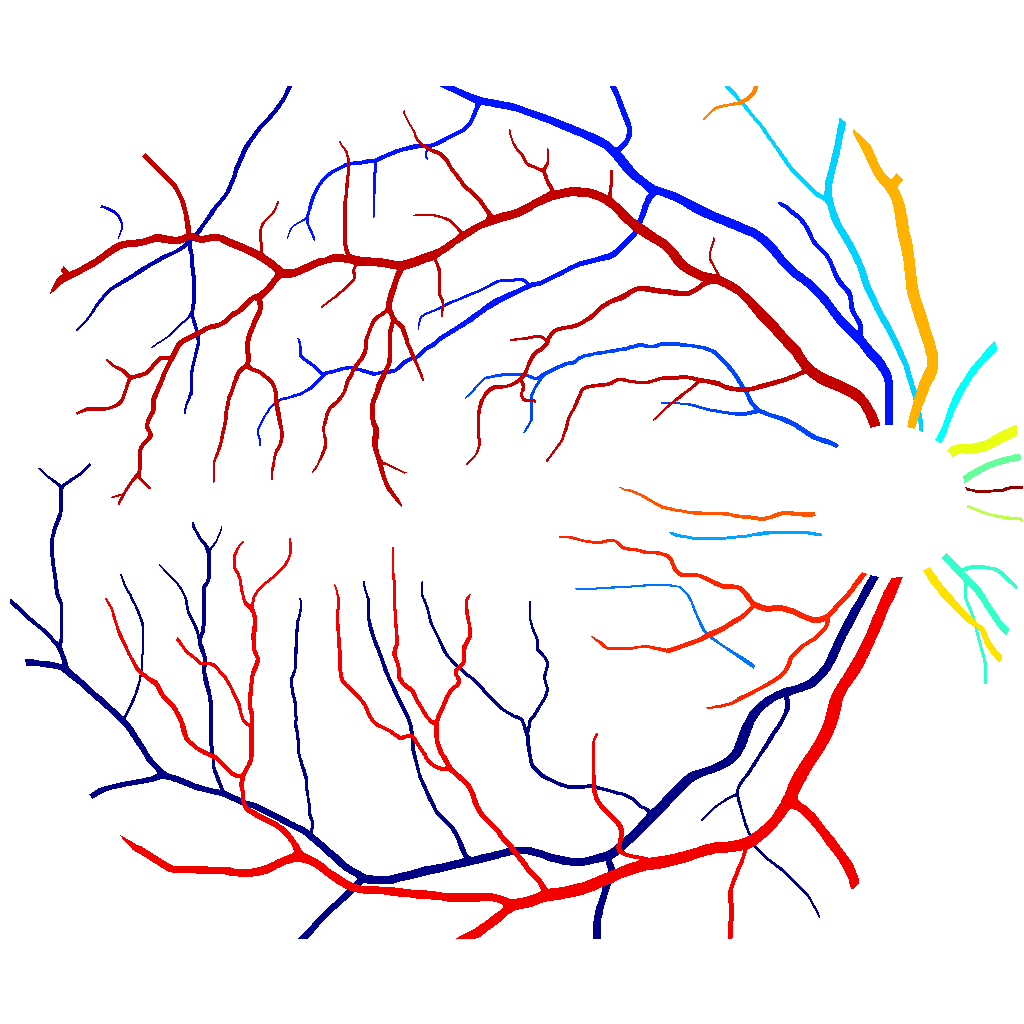}
         \caption{}
     \end{subfigure}
     \hfill
     \begin{subfigure}[b]{0.24\textwidth}
         \centering
         \includegraphics[width=\textwidth]{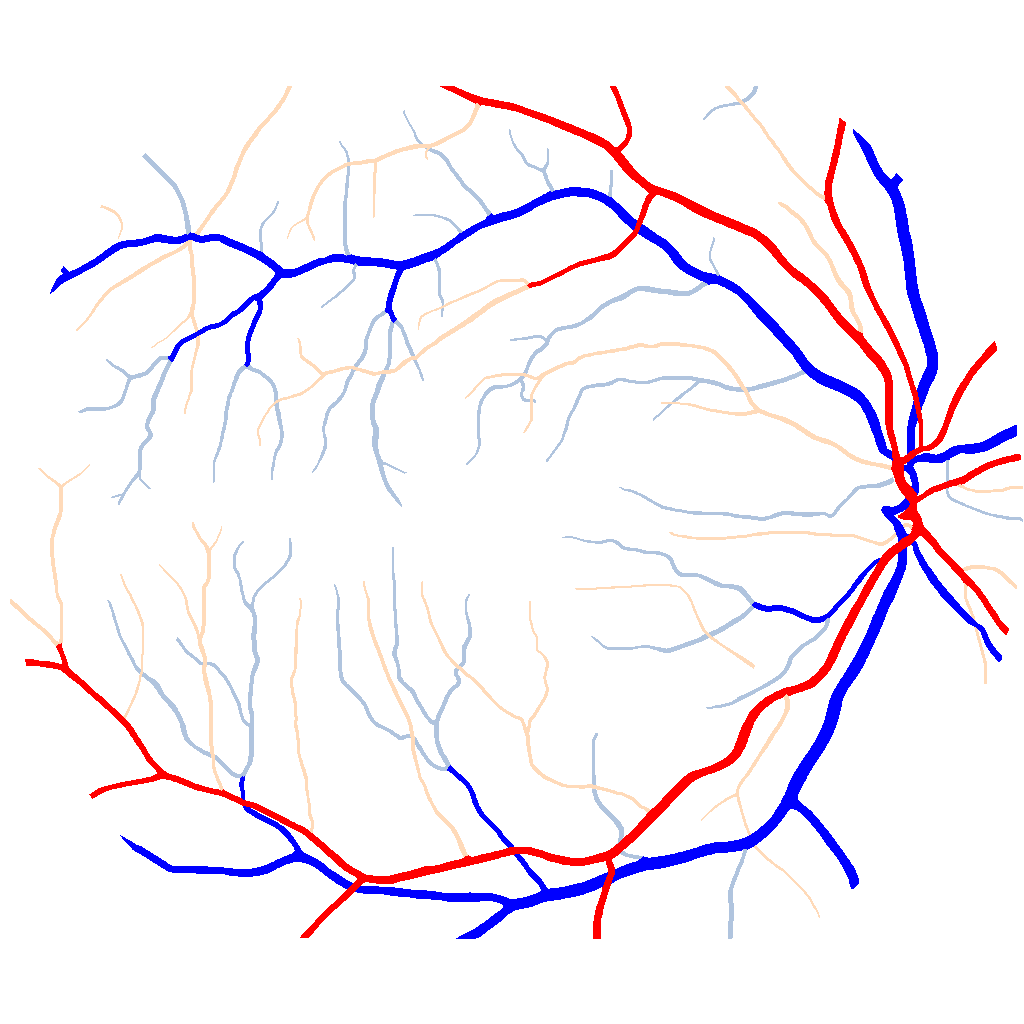}
         \caption{}
     \end{subfigure}
     \hfill
     \begin{subfigure}[b]{0.24\textwidth}
         \centering
         \includegraphics[width=\textwidth]{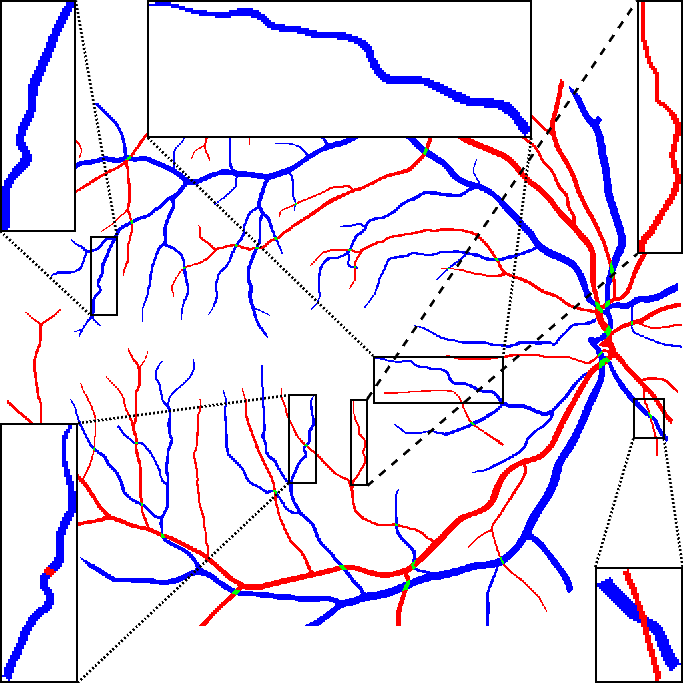}
         \caption{}
     \end{subfigure}
        \caption{An overview of annotations in RETA Benchmark. (a) original colour fundus image, (b)binary blood vessel mask, (c) artery/vein mask (Red: arterial vessel pixel, Blue: venous vessel pixel, Green: junctions of arteries and veins), (d) vessel skeletons(vessel centreline image is morphologically dilated with a 1 pixel disk-shaped structuring element for a better visualization), (e) vascular bifurcations (red and blue points represent arterial and venous bifurcations respectively) superimposed on greyscale image of (a), (f) vascular trees without OD region (every tree is encoded in an unique colour), (g) thick (Blue \& Red) and thin (Lighter Blue \& Red) vessel labels based on vessel calibre, (h) labelled vascular abnormalities highlighted on A/V mask (c). Disease class of lower right bounding box is arteriovenous nicking and class of other bounding boxes is vascular tortuosity.}
        \label{fig:1}
\end{figure}

In this work, we built a novel benchmark RETA for REtinal vascular Tree Analysis according to benchmark construction protocols\cite{kauppi2013constructing}. RETA contains 81 images inherited from the first subset of IDRiD dataset \cite{porwal2020idrid} and pixel-level blood vessel masks. A self-developed MATLAB-based interactive tool named as Computer Aided Retinal Labelling (CARL) is used for vessel annotation and visualization. The designed annotation approach consists of pixel-level, structure-level and network-level stages. Figure \ref{fig:1} (a) and (b) show a fundus image and corresponding labelled vessel mask. The intermediate outputs from different annotation stages are shown in Figure \ref{fig:1} (c)-(h). The potential reuse value for our RETA benchmark includes
\begin{itemize}
    \item Develop and evaluate automated approaches on blood vessel segmentation. The presence of different kinds of diabetic lesions like microaneurysms, soft exudates, hard exudates and hemorrhages in RETA colour fundus images makes segmentation a more challenging task\cite{mookiah2020review}.
    \item Build multi-task retinal image analysis models. Annotations of OD and lesions are publicly available in IDRiD dataset. Develop segmentation models to learn geographic relationship between vessel and other retinal objects could be an interesting research topic.
    \item Cross-modality tubular structure segmentation. Adapt feature domain from colour fundus photography to other medical imaging modalities, like optical coherence tomography angiography and magnetic resonance angiography , and segment vascular structures.
    \item Analyse retinal vascular trees. Track A/V trees with a correct topology. Measure topological and geometrical features of traced trees, such as branchpoint density, fractal dimension, calibre and tortuosity.
\end{itemize}

\section*{Methods}
\begin{figure}[ht]
\centering
\includegraphics[width=\linewidth]{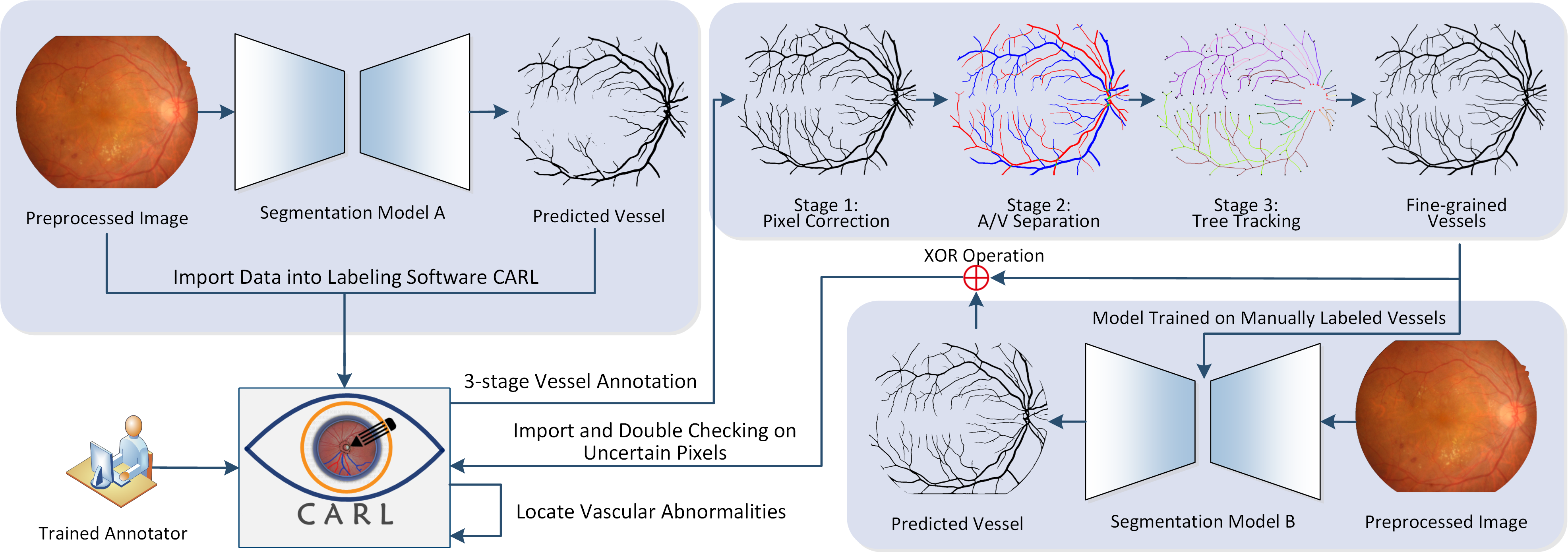}
\caption{Proposed workflow of generating fine-grained vessel annotation. Segmentation model A and B are automatic binary vessel segmentation models to predict vessel pixels. Stage 1 is pixel-level manual annotation on raw vessel segmentation from model A. Stage 2 is structure-level artery/vein segment identification. Stage 3 is used to validate A/V annotations from Stage 2 by means of network-level analysis. We tracked every tree starting from OD boundary. Different trees are encoded in unique colours with highlighted landmark points. Red rectangles are starting points. We further double-checked labels on non-consensual pixels between predictions from model B and manual annotations. Finally, we scanned labelled vessel pixels to localize vascular abnormalities. }
\label{fig:2}
\end{figure}

A semi-automated method is designed to build pixel-level vessel masks from coarse to fine. Figure \ref{fig:2} shows an overview of proposed method. Two vessel segmentation models are used to automatically generate vessel segmented images. Manual validation is mainly conducted on CARL software aiming to correct vessel pixel labels. In particular, the top right block of Figure \ref{fig:2} is a 3-stage vessel labelling strategy to label vessel pixel in fundus image. The first stage is pixel-level annotation. We started extensive manual vessel correction on computer predicted results from segmentation model A. Then structure-level segment labelling was performed to classify every vessel segment into artery or vein. In the third stage, we validated annotation quality in a network-level approach. The last step is to disambiguate vessel pixel labels between manual labelled mask and automated mask predicted from segmentation model B. The proposed method is able to control inter-annotator variability and intra-annotator variability because trained annotator performs on multi-stage annotation and subsequent label disambiguation. 

\subsection*{Image Acquisition.} The original colour fundus images are from task A in IDRiD Challenge. It contains 54 images for training and 27 images for testing. We resized original images from 4288$\times$2848 into 1024$\times$1024 dimensions with black image background cropped. Image down-sampling aims to reduce annotation time but keep small lesion like microaneurysms still visible in image. Human labeled lesion in original images will be invisible in our down-sampled images if its area is less than 15 pixels. The percentage of lost tiny microaneurysms and hard exudates are 2 out of 3,497 and 18 out of 11,642 respectively. Contrast enhanced images could assist vessel boundary delineation and tiny vessel recognition. In this work, the selected enhancement methods are contrast-limited adaptive histogram equalization (CLAHE) and local contrast enhancement (LCE) \cite{hemelings2019artery} method. The image transformation and enhancement algorithms are available in attached Codes. 

\subsection*{Automated Vessel Segmentation.} Pixel-level vessel labelling from scratch is an arduous task. Our approach is performing manual correction on pre-segmented vessel images. VGAN is a state-of-the-art method for vessel segmentation \cite{son2017retinal}. We predicted vessel pixels with probabilities ranged in [0,1] by a model trained on DRIVE \cite{staal2004ridge} dataset. The input fundus image is resized to 640$\times$640 and the output soft prediction is then resized its dimension from 640$\times$640 to 1024$\times$1024. Binarized vessel images are extracted by Sauvola thresholding method. For every fundus image, we generated a corresponding formatted mat file (a data container format in MATLAB) containing original retinal image, enhanced retinal image and segmented binary vessel for further vessel annotation in CARL. The mat file generation code is available in attached Codes. 

\subsection*{Vessel Pixel Correction.} Misclassified pixels in binary vessel images include background pixel wrongly classified as vessel pixel (false positive pixel) and vessel pixel wrongly classified as background pixel (false negative pixel). In CARL, binary vessel image is superimposed on fundus image with a changeable alpha to control vessel transparency. The background fundus image is switchable between original and enhanced images. The designed pixel-level annotation tools can remove or add pixels from current vessel maps. The correction time depends on accuracy of vessel pre-segmentation and labelling skill of annotator. Only one trained annotator took this work in order to control inter-annotator bias.

\subsection*{Artery/Vein Separation.} We observed that there are more venous pixels than arterial pixels in an image. It is proved that average percentages of arterial and venous pixels are 41.91\% and 58.09\% respectively in labelled A/V masks. Our strategy is separating veins from binary vessel image then masking out labelled veins from binary vessel image. Image completion on arteriovenous crossovers strives to connect disconnected arterial vessels. The classification features in terms of five different aspects, colour, calibre, crossover, light reflex and topology, between arteries and veins are in Table \ref{tab:1}. It is worth mentioning that we strictly classify every pixel into three classes (artery,vein or crossing) that is different from some existing datasets (RITE\cite{hu2013automated},LES-AV\cite{orlando2018towards} and HRF-AV\cite{hemelings2019artery}). In their A/V masks, they added an uncertain class referring to vessel pixels with unsure labels. 

In CARL, we firstly chose “Vein Vessel” from “Vessel Types”. Binary vessel mask will overlay on colour fundus image. We zoomed into a crossover (usually connects 4 to 6 vessel segments), identified venous segments, disconnected and removed arterial segments. In this stage, we named this vessel class labeling as structure-level vessel annotation. Segments between two crossovers will share the same vessel class. We also categorized a crossover into artery over vein and vein over artery. Crossover type is crucial information for predicting vessel position from 2D fundus images not 3D optical coherence tomography scans \cite{kumagai2014three}. While labeling A/V labels, we paid special attention to vessel boundaries and crossovers where boundary delineation is usually hard to determine\cite{wang2019retinal}. 

\begin{table}[ht]
\centering
\begin{tabular}{|l|l|}
\hline
\textbf{Features} & \textbf{Differences between arteries and veins}                                                                                                                                            \\ \hline
Colour            & Central veins have darker colour than central arteries.                                                                                                                                    \\ \hline
Calibre           & Veins have wider diameters than adjacent arteries.                                                                                                                                         \\ \hline
Crossover         & \begin{tabular}[c]{@{}l@{}}Vessels with same labels hardly cross each other. Arteriovenous nicking \\ can only be seen when an arterial vessel crossing over a venous vessel.\end{tabular} \\ \hline
Light reflex      & Veins show a smaller central light reflex.                                                                                                                                                 \\ \hline
Topology          & Veins and arteries are usually alternate to each other near optic disk.                                                                                                                    \\ \hline
\end{tabular}
\caption{Classification of retinal arteries and veins in 2D fundus image}
\label{tab:1}
\end{table}

\subsection*{Vessel Tree Tracking.} We constructed vascular graph from A/V annotations and measured topological and geometrical properties of built graph. Graph vertex and edge refer to landmark point (bifurcation or terminal point) and a vessel segment that connects two landmark points in vessel skeleton image. We tracked all vascular trees in this graph and modified potential topological errors and geometrical abnormalities. 

\subsubsection*{Vessel Skeletonization.} Vessel calibre varies from 1 to 20 pixels in RETA benchmark. Skeletonization reduces thick or thin vessels to 1-pixel width representations. As one of tubular structures, retinal blood vessel can be presented as the envelope of a family of disks with continuously changing centre points and radii \cite{benmansour2011tubular}. A desirable vessel skeleton demands no spurs (spurious lines) on extracted centrelines. We implemented Lee’s skeletonization method to extract vessel skeletons without unwanted spurs on vessel boundaries or end points \cite{lee1994building}. However, it failed to detect accurate vascular centrelines at image border where partial cross-section line of vessel structure is outside the image. In this case, we had to manually correct these regions. We performed above approach on arterial mask and venous mask separately. 

\subsubsection*{Graph Construction and Visualization.} 
\begin{figure}[ht]
\centering
\includegraphics[width=0.55\linewidth]{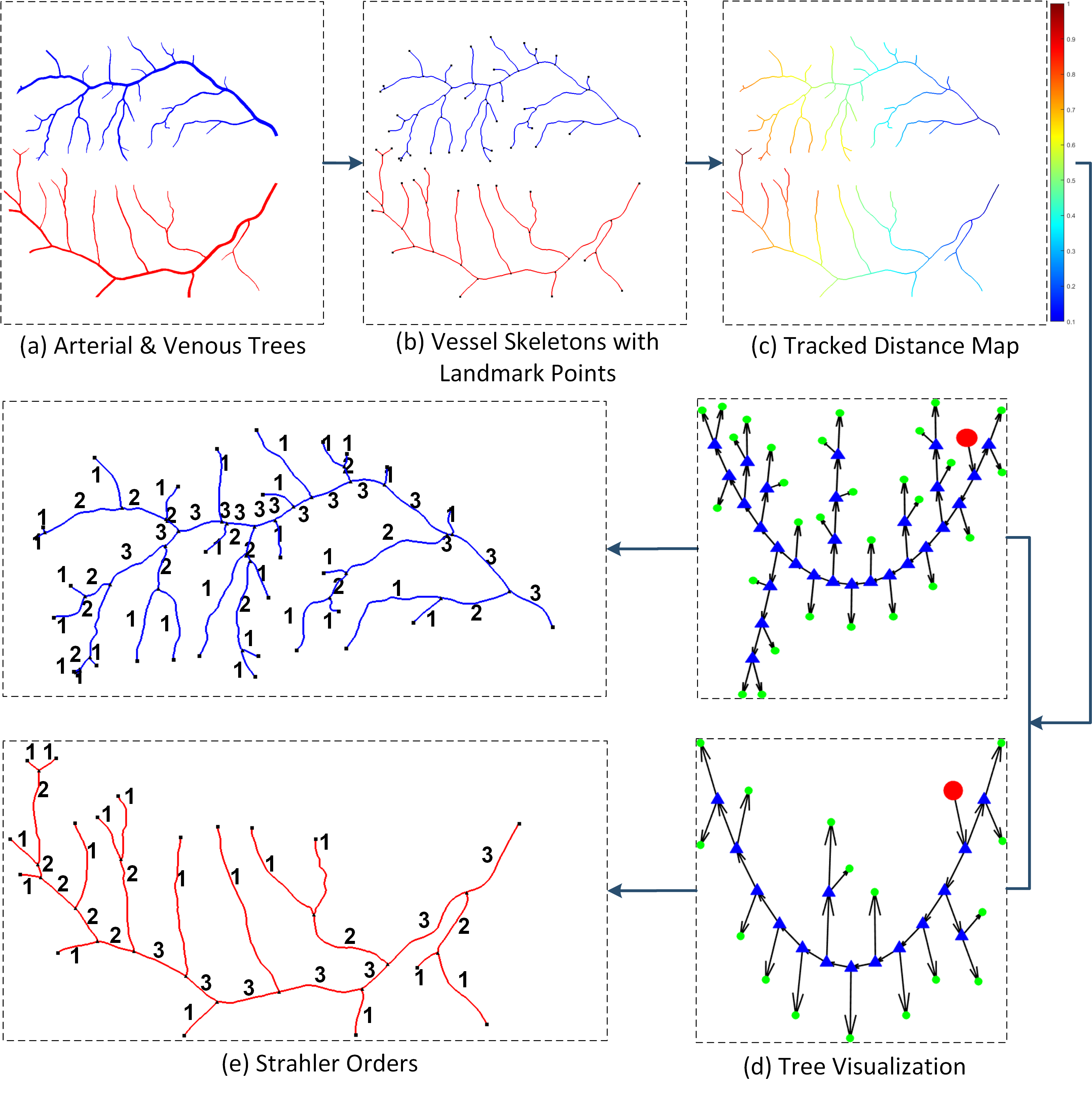}
\caption{An example of vascular tree tracking. (a) red and blue pixels indicate arterial and venous trees, (b) vessel skeleton image marked with landmark points in black, (c) normalized pixel distance away from rightmost pixel of vessel centreline image, (d) tracked vascular tree (Red dots, blue rectangles and green dots are starting, bifurcation and ending vertices respectively in graph), (e) Strahler number assigned to every vessel segment.}
\label{fig:3}
\end{figure}

We removed A/V skeleton pixels within OD region since OD contains intertwined vessels and vascular graph built inside OD might be unreliable. Then, we selected three different types of points from vessel centrelines as graph vertices. A 3*3 image filter is employed to count neighbours for every pixel in vessel skeleton image. Every skeleton pixel has 1 to 4 neighbours. We defined a pixel with at least three neighbours as bifurcation vertex. Starting and ending vertices are pixels with only two neighbours. The difference between them is starting vertex is usually close to OD boundary. The next step is building adjacency matrix for extracted vertices. We proposed a graph region growing method to track A/V skeleton image from starting vertices. Each vascular tree is visualized by TreeVis\cite{qiu2013treevis}. Figure \ref{fig:3}(d) shows tracking results of 2 different trees where red dots indicate the starting vertices.

\subsubsection*{Topological Abnormalities Detection.} A normal rooted arterial/venous vascular tree is a binary tree where a parent vertex only connects two child vertices without graph circles. We proposed bifurcation vertex validation and graph circle detection methods to detect topologically erroneous based on above assumption. For any bifurcation vertex, we computed the number of connected edges while labeling Strahler number to each edge. Image annotator validated abnormal bifurcation vertex with more than 3 connected edges. There is an exception when 2 vessel bifurcations are extremely close in geometrical space and graph construction method merges 2 bifurcation vertices into only 1. As for graph circle detection, we built a minimum spanning tree (MST) of each vascular tree (VT). Graph circle exists if there is any edge in VT but not in MST.

\subsubsection*{Geometrical Abnormalities Detection.} Three significant vessel geometrical properties are vessel calibre, bifurcation angle and tortuosity. Most vascular abnormalities are closely associated with these geometrical property changes. Diameter changes are visible in venous beading, arteriolar/venous narrowing, arteriovenous nicking and branch retinal vein occlusion. Venous loop is one of tortuosity patterns. We identified vessel segment mask whose skeleton connects 2 vertices in constructed graph. Vessel diameter estimation method\cite{bankhead2012fast} is applied to measuring diameters at all centreline pixels. We tracked two child edges at a bifurcation vertex and computed bifurcation angle between two child vessel skeletons in a 10-pixel searching area. Vessel tortuosity is also measured on vessel skeleton for all edges\cite{khansari2017method}. We assumed vessel geometrical abnormalities meet any of the following condition: 
\begin{itemize}
\item Focal diameter variation $>3$ pixels.
\item Average calibre of a parent segment $<$ average calibre of any child segment.
\item Bifurcation angle $>150$\textdegree.
\item Vessel tortuosity index $>5$.
\end{itemize}
Image annotator is required to review and update vessel annotation for all abnormal segments. Moreover, we generated thick (segment mean calibre $>$ 5 pixels) and thin vessel (segment mean calibre $<$ 5 pixels) masks and abnormal vessel locations based on calculated geometrical features.

\subsection*{Disambiguation of pixel labels.}
 In annotated vessels, noisy pixels are usually introduced at vessel boundaries by human annotators. Detecting and fixing these noisy pixels can improve vessel annotation quality. The noise identification method we proposed is to train a vessel segmentor (another VGAN model) on labelled vessel image and identify potential incorrect pixels based on the predictions of this segmentor. This is a popular label cleaning approach based on the assumption that misclassified pixels can be probably noisy pixels \cite{karimi2020deep}. Then, we compared training labels with predicted labels and categorized non-consensus labels into two noise types, unlabelled and mislabelled. We further checked these noisy labels by experienced annotator in CARL software. A final vessel annotation is obtained after removing mislabelled pixels and adding unlabelled pixels. This operation could control intra-annotator variability. 

\section*{Data Records}
Users can access RETA benchmark by visiting our website at \href{https://www.reta-benchmark.org}{https://www.reta-benchmark.org}. We released a shared folder named as RETA includes labelled vessel masks and additional data. It contains five subfolders, namely “codes”, “images”, “mats”, “models” and “software”. Annotations are saved in “images” and “mats” folders. Code Availability section will introduce contents within “codes”, “models” and “software” folders.

In the “images” folder, we divided all the images into “train” and “test” subfolders. In each subfolder, colour fundus images, fundus masks and vessel masks are stored in “img”, “mask”, “vessel” folders respectively. Colour fundus images are resized images not original images of IDRiD Challenge but we remain file names unchanged. A fundus mask refers to a binary image indicating all captured retinal pixels in camera field of view (FOV). Fundus and vessel masks are saved in PNG format and named as “*\_mask.png” and “*\_vessel.png” where * indicates file name of corresponding fundus image. There are 54 images for training and 27 images for testing same as in IDRiD Challenge.

\begin{table}[ht]
\centering

\begin{tabular}{|l|l|l|}
\hline
\textbf{Structure}           & \textbf{Field} & \textbf{Definition}                                                                                                                                   \\ \hline
\multirow{9}{*}{MAT}         & I              & \begin{tabular}[c]{@{}l@{}}A 3-D integer array denotes colour fundus image (may be resized if original \\ fundus image is too large)\end{tabular}      \\ \cline{2-3} 
                             & I\_cropped     & \begin{tabular}[c]{@{}l@{}}A 3-D integer array denotes fundus image cropped from I (see pos\_data \\ structure for cropping parameters).\end{tabular} \\ \cline{2-3} 
                             & enhG\_cropped  & A 3-D integer array denotes CLAHE enhanced image of I\_cropped.                                                                                       \\ \cline{2-3} 
                             & enhC\_cropped  & A 3-D integer array denotes LCE enhanced image of I\_cropped.                                                                                        \\ \cline{2-3} 
                             & mask\_white\_o & A 2-D logical array denotes fundus mask of I.                                                                                                         \\ \cline{2-3} 
                             & mask\_white    & A 2-D logical array denotes fundus mask of I\_cropped.                                                                                                \\ \cline{2-3} 
                             & dim\_change      & A structure array contains indicators of image dimension changes.                                                                                     \\ \cline{2-3} 
                             & pos\_data      & A structure array contains image cropping parameters.                                                                                                 \\ \cline{2-3} 
                             & annotations    & A structure array contains all kinds of labelled pixels.                                                                                              \\ \hline
\multirow{2}{*}{dim\_change}   & row            & \begin{tabular}[c]{@{}l@{}}Boolean 1 and 0 denote height of an original fundus image is odd or even \\  respectively. If 1, the first row of the image will be removed.\end{tabular}                      \\ \cline{2-3} 
                             & col            & \begin{tabular}[c]{@{}l@{}}Boolean 1 and 0 denote width of an original fundus image is odd or even \\  respectively. If 1, the first column of the image will be removed.\end{tabular}                       \\ \hline
\multirow{9}{*}{pos\_data}   & ori\_row       & An integer denotes height of I.                                                                                                                       \\ \cline{2-3} 
                             & ori\_col       & An integer denotes width of I.                                                                                                                        \\ \cline{2-3} 
                             & cropped\_row   & An integer denotes image height after image cropping operation.                                                                                       \\ \cline{2-3} 
                             & cropped\_col   & An integer denotes image width after image cropping operation.                                                                                        \\ \cline{2-3} 
                             & cropped\_left  & An integer array specifies upper-left location of cropping bounding box.                                                                           \\ \cline{2-3} 
                             & cropped\_right & An integer array specifies bottom-right location of cropping bounding box.                                                                         \\ \cline{2-3} 
                             & extension      & \begin{tabular}[c]{@{}l@{}}An integer array specifies added background pixels if contour of image FOV is not circular.\end{tabular}                 \\ \cline{2-3} 
                             & row            & An integer denotes height of original fundus image after dimension change.                                                                            \\ \cline{2-3} 
                             & col            & An integer denotes width of original fundus image after dimension change.                                                                             \\ \hline
\multirow{4}{*}{annotations} & label          & A string denotes annotation type. In this work, ‘BV’ indicates blood vessel.                                                                                        \\ \cline{2-3} 
                             & category       & A numerical value corresponding to given label. Category of 'BV' is 4 defined in CARL.                                                                            \\ \cline{2-3} 
                             & sure\_inds     & Linear indices of the pixels belonging to given label.                                                                                                \\ \cline{2-3} 
                             & unsure\_inds   & Linear indices of the pixels not belonging to given label.                                                                                             \\ \hline
\end{tabular}
\caption{Data structure in custom mat file for CARL software. A mat file contains four structure arrays named as "MAT", "dim\_change", "pos\_data" and "annotations".}
\label{tab:2}
\end{table}

In the “mats” folder, we provided custom MATLAB mat files generated in this work. Users could import these files into CARL and visualize zoomed tiny or complex vessel structures. All mat files are named as “*\_labeled.mat” and * indicates fundus image name. We stored all the parameters regarding to an image into a structure array “MAT”. Table \ref{tab:2} shows each structure field and its definition in the mat file. 

\section*{Technical Validation}
In the process of labelling binary vessel mask, additional types of vessel annotations are generated at different annotation stages in order to improve the quality of binary vessel annotation. The extra annotations include arterial/venous mask, skeletons, bifurcations and trees shown in Figure \ref{fig:1}. Accuracy of arterial/venous pixel-level annotation is of greater importance because it could ensure reliable annotations of vessel skeletons, bifurcations and trees. Also, a completed and reliable binary vessel masks depends on annotation quality of arterial and venous vessels. A binary vessel is a combination of arterial and venous masks. In this section, we focused on validating the reliability of final binary vessel masks by means of subjective and objective approaches. 

\begin{table}[ht]
\begin{tabular}{@{}llll@{}}
\toprule
\textbf{Score} & \textbf{Quality} & \textbf{Grading Criteria}                                                                                                                                                                                                                                                                                                           & \textbf{Example} \\ \midrule

1     & Unacceptable   & \begin{tabular}[c]{@{}l@{}}Any one of the following conditions:\\ (1) 2 disconnected vessel segments or more;\\ (2) Overlapping degree is smaller than 50\%;[\checkmark] \\ (3) More than 3 unlabeled vessel segments;[\checkmark]\\ (4) Unsmooth vessel edges and quite unclear boundary \\     delineation at vascular junctions.[\checkmark]\end{tabular}                         & \begin{tabular}[c]{@{}l@{}}\includegraphics[width=5.28cm, height=2.65cm]{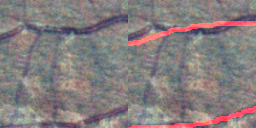} \end{tabular}        \\
2     & Poor        & \begin{tabular}[c]{@{}l@{}}Any one of the following conditions:\\ (1) Only 1 disconnected vessel segment;\\ (2) Overlapping degree is close to 75\%;\\ (3) 1-3 unlabeled vessel segments;\\ (4) There are many unsmooth pixels/spurs at vessel edges \\     and unclear boundary delineation at vascular junctions.[\checkmark]\end{tabular}    &  \begin{tabular}[c]{@{}l@{}}\includegraphics[width=5.28cm, height=2.65cm]{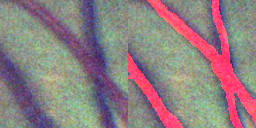} \end{tabular}       \\
3     & Fair       & \begin{tabular}[c]{@{}l@{}}All the following conditions:\\ (1) Connected vessel segments;\\ (2) Overlapping degree is close to 90\%;\\ (3) Hardly see any unlabeled vessel segment;\\ (4) There are some unsmooth pixels/spurs at vessel edges \\     and unclear boundary delineation at vascular junctions.\end{tabular} &   \begin{tabular}[c]{@{}l@{}} \includegraphics[width=5.28cm, height=2.65cm]{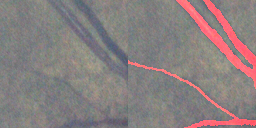} \end{tabular}     \\
4     & Good       & \begin{tabular}[c]{@{}l@{}}All the following conditions:\\ (1) Connected vessel segments;\\ (2) Overlapping degree is close to 95\%;\\ (3) No vessel segment unlabeled;\\ (4) There are some unsmooth pixels/spurs at vessel edges \\     but clear boundary delineation at vascular junctions.\end{tabular}         & \begin{tabular}[c]{@{}l@{}}\includegraphics[width=5.28cm, height=2.65cm]{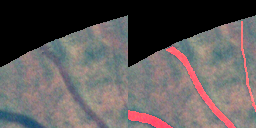} \end{tabular}       \\
5     & Excellent  & \begin{tabular}[c]{@{}l@{}}All the following conditions:\\ (1) Connected vessel segments;\\ (2) Overlapping degree is close to 100\%;\\ (3) No vessel segment unlabeled;\\ (4) Smooth vessel edges and clear boundary delineation \\     at vascular junctions.\end{tabular}                                                  &  \begin{tabular}[c]{@{}l@{}}\includegraphics[width=5.28cm, height=2.65cm]{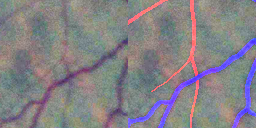} \end{tabular}      \\ \bottomrule
\end{tabular}
\caption{Grading scale for subjective vessel annotation quality. For predicting the percentage \% of overlapping degree, it is subjectively estimated by human evaluator. In each pair of exemplar images, the left image is a $128 \times 128$ LCE enhanced image patch. The right one shows labeled vessel pixels covering over the left image. In unacceptable and poor quality categories, there is [\checkmark] logo if the condition is met for the exemplar vessel annotation.}
\label{tab:3}
\end{table}

The most common method for evaluating quality of image segmentation mask is subjective evaluation. Nevertheless, it is tedious and time-consuming for human evaluators. In objective evaluation, unsupervised methods focus on designing quantitative metrics such as object shape in segmented mask\cite{zhang2008image}. But appropriate measures are hard to define for complicated retinal vascular trees. A similar regression model between artificial degraded and human-labelled vessel masks is trained to predict quality of any input vessel segmented mask\cite{galdran2018no}. However, model accuracy is highly dependent on quality of reference images in the training set.

\subsection*{Subjective Quality Evaluation.} 
We proposed four heuristic metrics for subjective vessel quality evaluation. They are segment connectivity, overlapping degree, mislabeled quantity and edge smoothness. Segment connectivity demands pixels of a vessel segment should be connected and not broken. Overlapping degree measures the overlap degree of marked and ground-truth blood vessel pixels. It is a region based metric similar to Dice Coefficient. Mislabeled quantity refer to the number of unlabeled ground-truth and labeled artificial vessel segments. Edge smoothness is based on the assumption that normal blood vessel has smooth boundaries like a tubular structure. Jagged vessel edges are caused by unskilled image annotators. Table \ref{tab:3} is a proposed 5-level grading scale integrating above metrics for vessel annotation quality evaluation. 

One trained evaluator graded vessel masks in RETA dataset and 11 publicly available datasets (DRIVE, STARE\cite{hoover2000locating}, ARIA\cite{farnell2008enhancement}, CHASE\_DB1\cite{fraz2012ensemble}, HRF\cite{budai2013robust}, DualModal2019\cite{zhang2019simultaneous}, DRHAGIS\cite{holm2017dr}, UoA-DR\cite{chalakkal2017comparative}, LES-AV, ORVS\cite{sarhan2021transfer} and AFIO\cite{akram2020data}). We randomly selected 5 images from each dataset and made mat files using the approach described in Image Acquisition section. The evaluator imported generated mat files into CARL and graded quality of vessel masks with the help of an embedded toolkit, annotation quality rater (AQR). AQR will split a 2D image of 1024$\times$1024 pixels into image patches of 128$\times$128 pixels. The evaluator assigned a score between 0 to 5 for every image patch. 0 for no visible vessel in current image patch. The grading process is available in the "software" folder of published dataset. Then, we computed average quality score for all the image patches that scored as 1 to 5 within a dataset. Table \ref{tab:4} is the results (mean±std) for all 12 image sets. Well-known DRIVE and STARE datasets suffer from vessel boundary smoothness and achieve lower quality scores. One possible reason is limited hardware and software facilities for retinal vessel annotation task nearly 20 years ago. With the help of CARL software, RETA dataset shows the highest vessel annotation quality and might be a novel benchmark in this research field.

\begin{table}[ht]
\centering
\begin{tabular}{|l|l|l|}
\hline
\textbf{Rank} & \textbf{Dataset} & \textbf{Quality Score} \\ \hline
1             & RETA             & 3.0880±1.4306          \\ \hline
2             & DualModal2019    & 2.4021±1.5130          \\ \hline
3             & DRHAGIS          & 2.2792±1.4632          \\ \hline
4             & LES-AV           & 2.2625±1.6988          \\ \hline
5             & CHASE\_DB1         & 2.2035±1.2230          \\ \hline
6             & HRF              & 2.1925±2.2646          \\ \hline
7             & AFIO             & 2.1715±0.9754          \\ \hline
8             & ORVS             & 1.9699±1.6102          \\ \hline
9             & ARIA             & 1.7402±1.0803          \\ \hline
10            & STARE            & 1.7323±1.0928          \\ \hline
11            & DRIVE            & 1.6630±1.2630          \\ \hline
12            & UoA-DR           & 1.1187±0.6480          \\ \hline
\end{tabular}
\caption{Subjective vessel annotation quality assessment of RETA and 11 public datasets.}
\label{tab:4}
\end{table}

\subsection*{Objective Quality Validation.} 
In this section, we applied four validation methods to objectively evaluating labeled vessel masks. Hole detection and connected-component analysis are two common image processing techniques. Human observers are in the loop of validation process to classify detected pixels into noise or vessel category. Furthermore, we used fractal dimension (FD) of retinal vasculature as a clinical metric for evaluating our labeled vessel images since change of FD is associated with disease progression\cite{yu2021fractal,thomas2014measurement}.  Finally, we trained vessel segmentation models on RETA and two public database and analyzed the model optimization process. Our assumption is computer learns harder and longer to a optimal point if training set contains more noisy labels.

\subsubsection*{Hole detection.} A hole is a set of black pixels inside a white object in binary images. Holes in labelled vessel masks will lead to inaccurate vessel skeletons. Hole detection can locate not only vascular loops but unlabeled pixels inside a vessel as well. In RETA, holes are hardly seen in arterial/venous binary masks except twisted vessels inside OD or image with venous loops. A flood-fill operation on binary image is used to find unexpected holes. There are 6 holes marked as venous loop and only 2 holes detected within OD in all 81 images.

\subsubsection*{Connected-component analysis.} Disconnected small connected components (CC) are prone to be erroneous annotations. Small CCs could be mislabelled background pixels or isolated set of pixels disconnecting from large CCs. Connected-component analysis can detect small vessel segments. We performed this method on binary arterial/venous masks respectively. All CCs with area smaller than 100 pixels are double-checked. We identified 36 small CCs (11 for arteries and 25 for veins) and none of them belongs to background pixels.

\subsubsection*{Fractal dimension analysis.}
Fractal dimension analysis is a mathematical method to describe geometric complexity of retinal vascular trees. We used box-counting dimension\cite{liebovitch1989fast} to estimate FD on manual-labeled vessel masks. Two trained human readers graded diabetic retinopathy (DR) and diabetic macular edema (DME) levels for all colour fundus images in RETA according to the International Clinical Diabetic Retinopathy Scale\cite{wilkinson2003proposed}. The FD measured areas include entire FOV region and standardized macula region (radius of macula is 0.6 times the distance of macula center to OD center). DME is closely related to vascular changes within macula region.

\begin{figure}[ht]
     \centering
     \begin{subfigure}[b]{0.23\textwidth}
         \centering
         \includegraphics[width=\textwidth]{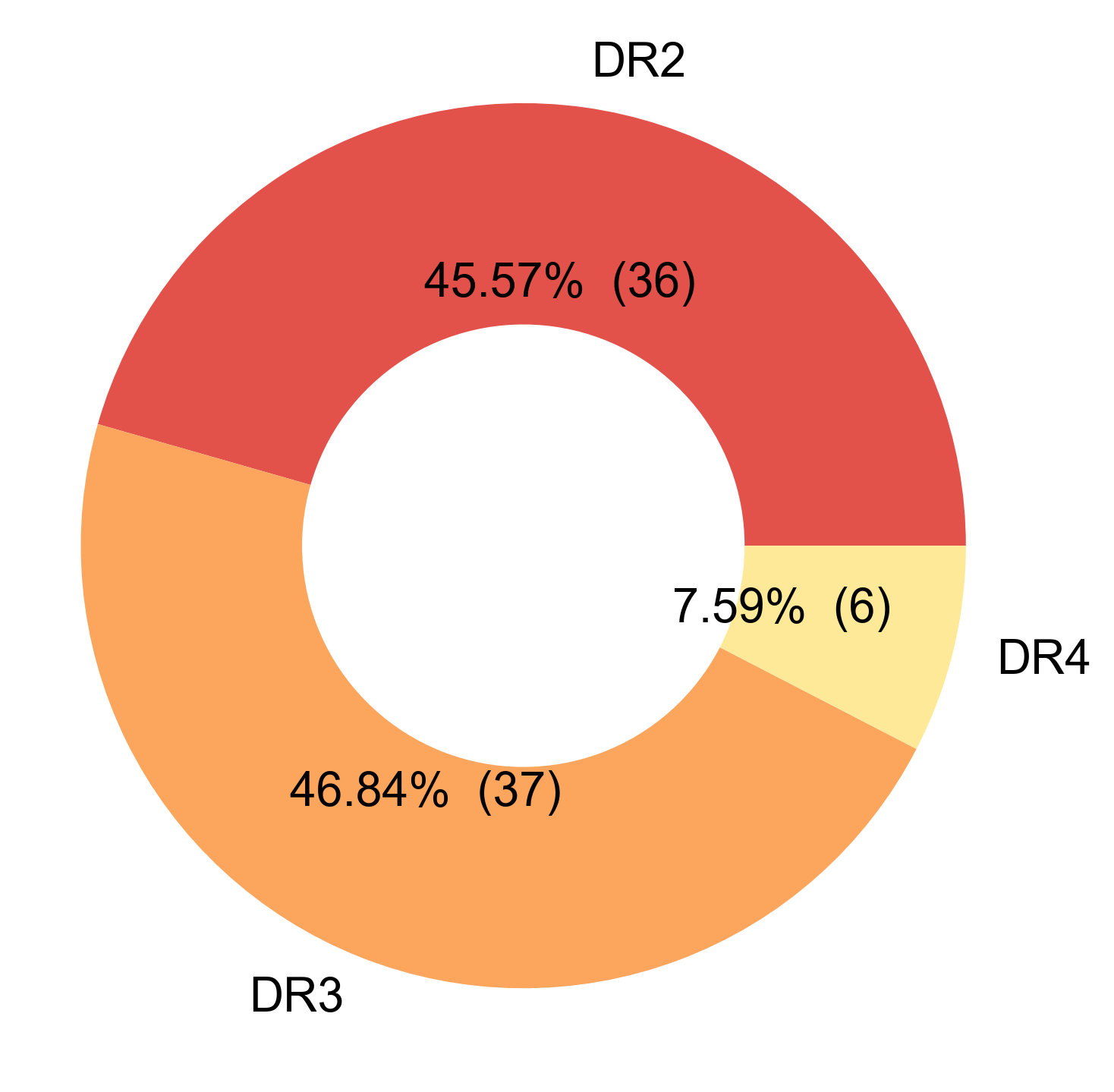}
     \end{subfigure}
     \hfill
     \begin{subfigure}[b]{0.37\textwidth}
         \centering
         \includegraphics[width=\textwidth]{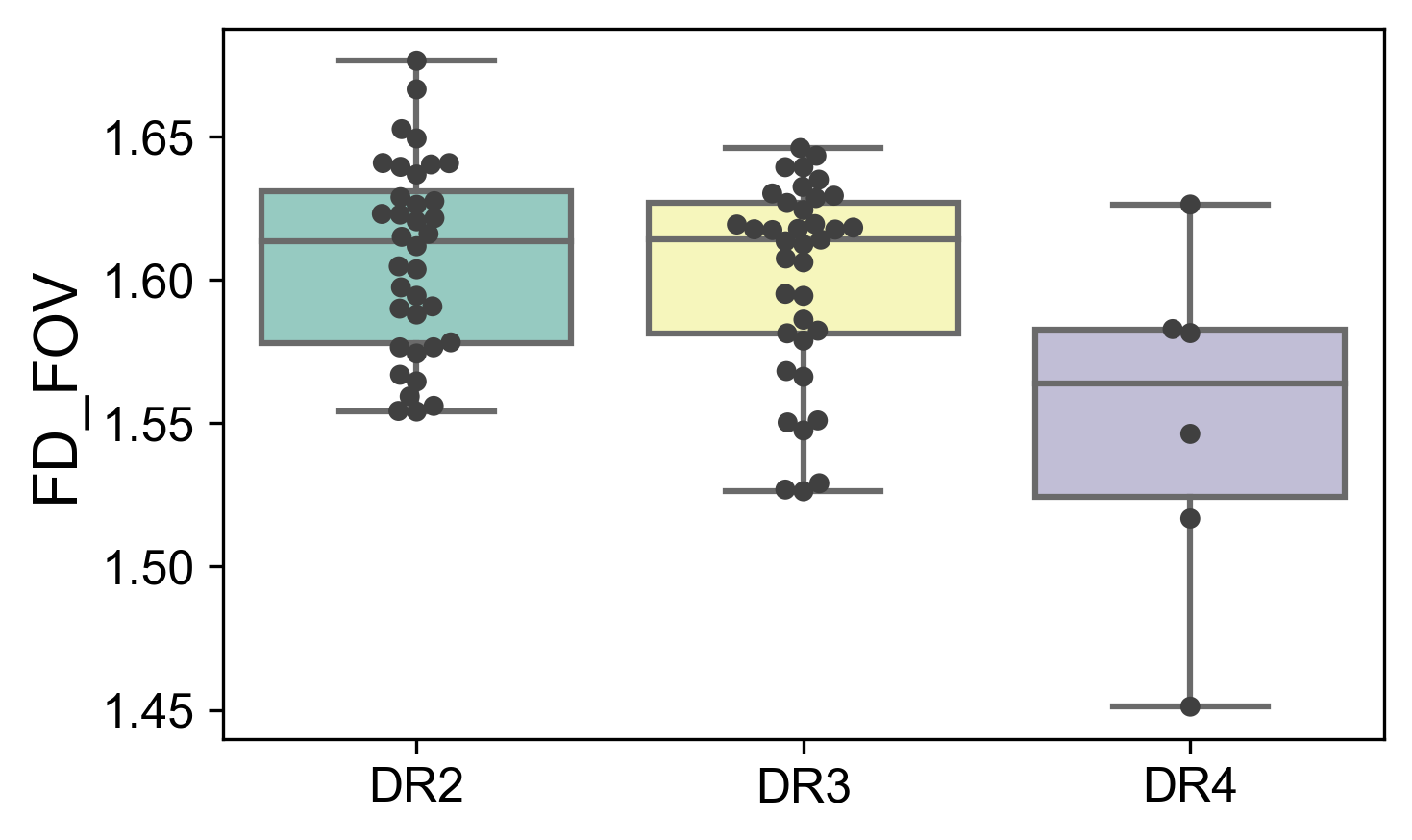}
     \end{subfigure}
     \hfill
     \begin{subfigure}[b]{0.37\textwidth}
         \centering
         \includegraphics[width=\textwidth]{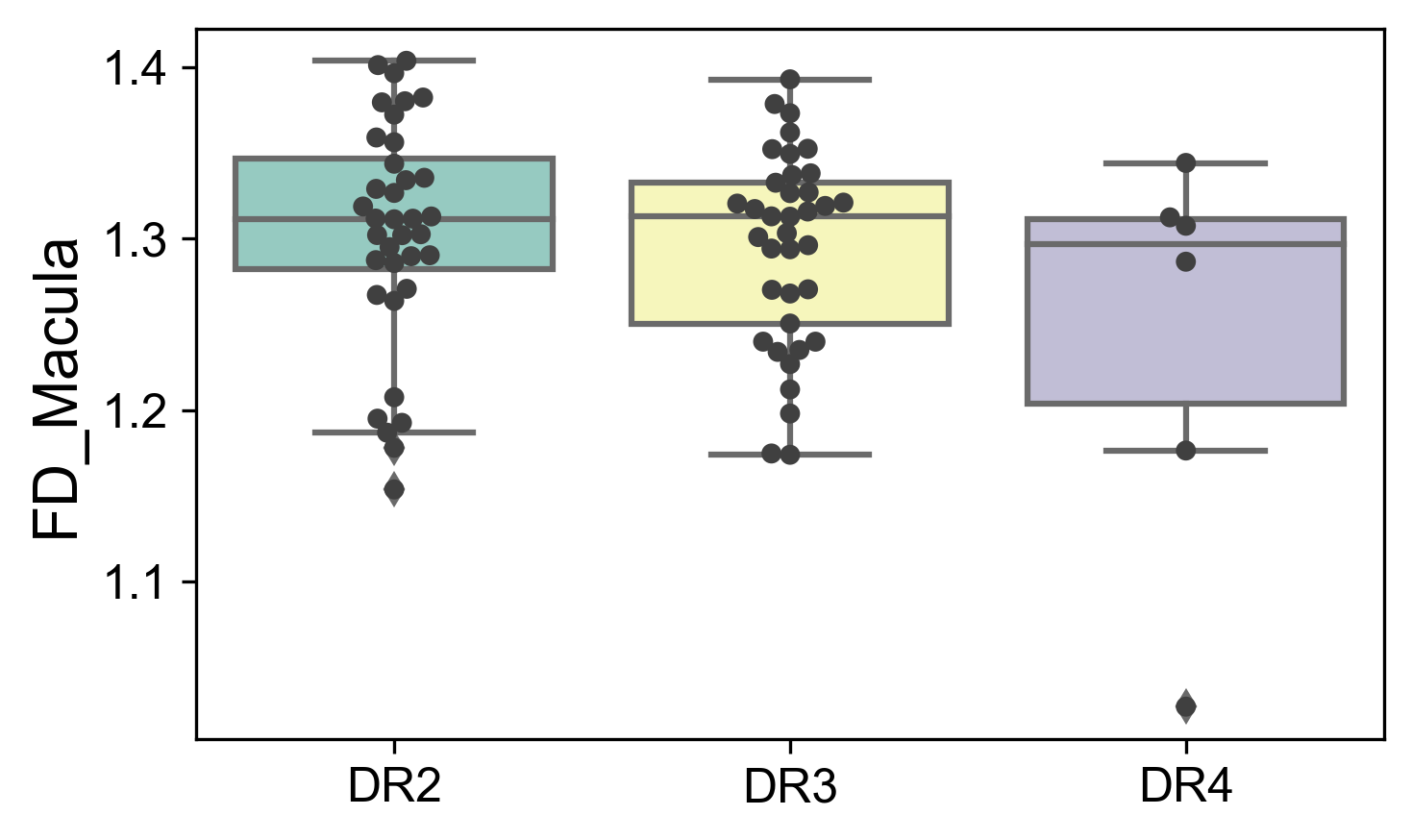}
     \end{subfigure}
     
     \begin{subfigure}[b]{0.23\textwidth}
         \centering
         \includegraphics[width=\textwidth]{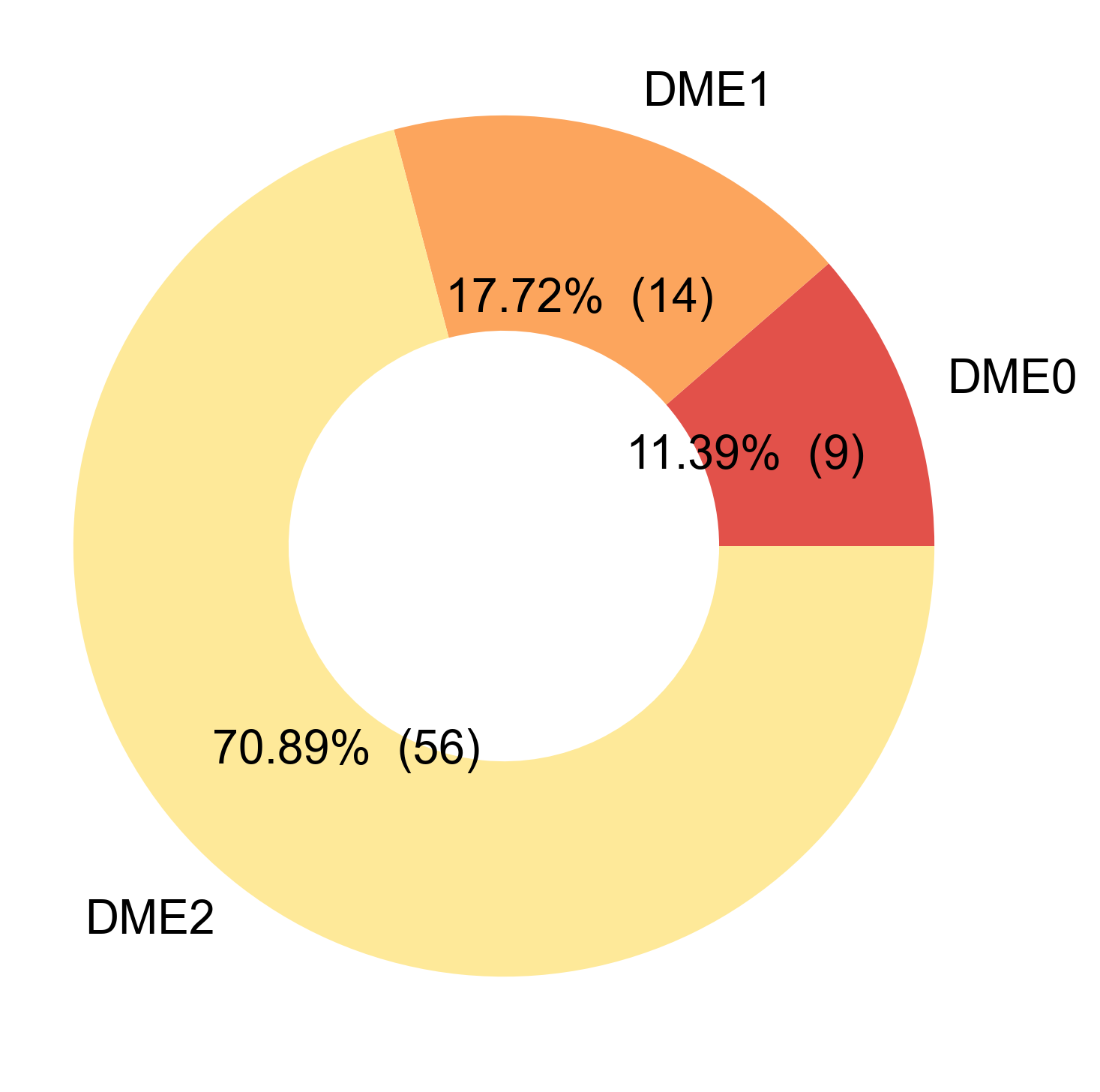}
     \end{subfigure}
     \hfill
     \begin{subfigure}[b]{0.37\textwidth}
         \centering
         \includegraphics[width=\textwidth]{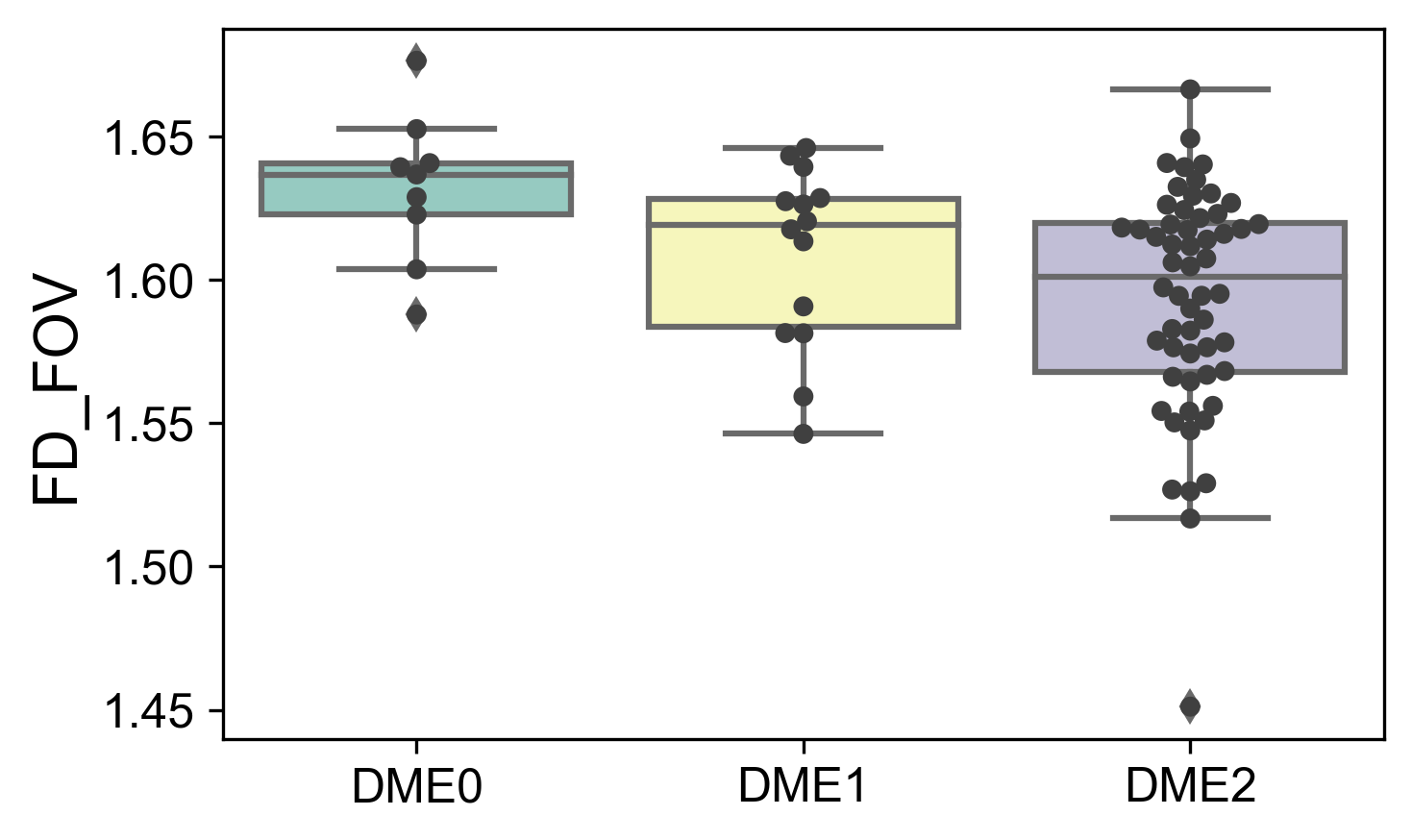}
     \end{subfigure}
     \hfill
     \begin{subfigure}[b]{0.37\textwidth}
         \centering
         \includegraphics[width=\textwidth]{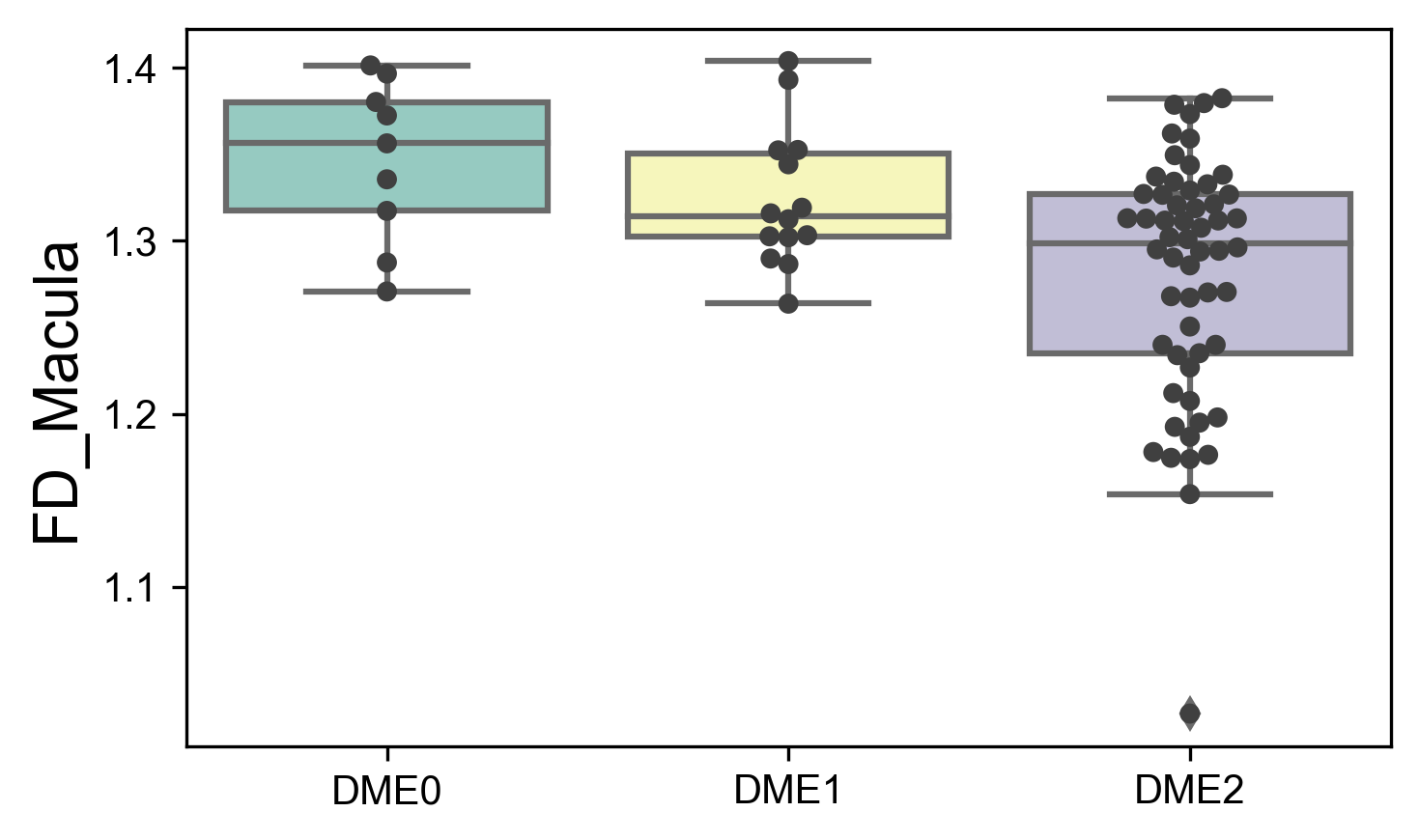}
     \end{subfigure}
        \caption{Study group distribution and box plots of fractal dimensions. The second and third columns show FD of entire FOV and macula region in different groups.}
        \label{fig:4}
\end{figure}

Two images (named as "IDRiD\_04.jpg" and "IDRiD\_45.jpg") are excluded from this study since they are OD-centered fundus images. Doughnut charts in Figure \ref{fig:4} are image distribution of different DR (DR2: moderate non-proliferative DR, DR3: severe non-proliferative DR and DR4: proliferative DR) and DME (DME0: normal, DME1: non-clinically significant DME and DME2: clinically significant DME) groups. The box plots in Figure \ref{fig:4} showed mean FD and 95\% confidence intervals. We only compared FD of DR2 and DR3 groups since sample size of DR4 is not big enough. T-test showed there was no significant difference between them in entire FOV ($p=0.3892$) and macula region ($p=0.5520$). For DME, FD of DME0 and DME2 was significant ($p<0.05$) in both regions but FD of DME1 and DME2 was only significant ($p<0.05$) in macula region. The progression of DME severity was found to be associated with lower FD. This finding was consistent with previous clinical studies\cite{thomas2014measurement}.

\subsubsection*{Convergence process of model optimization.}
\begin{figure}[h]
\centering
\includegraphics[width=0.68\linewidth]{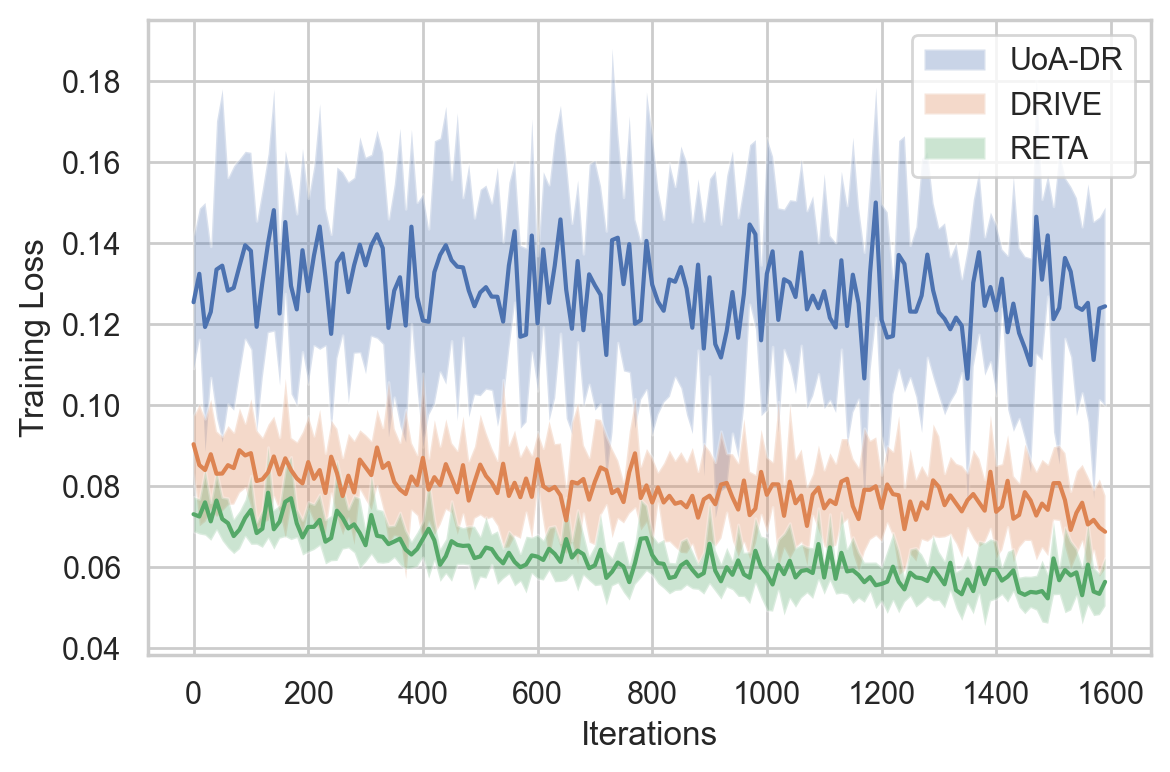}
\caption{The training loss over iteration curve of the first epoch learning to segment retinal vessels on UoA-DR, DRIVE and RETA datasets.}
\label{fig:5}
\end{figure}

Optimization algorithm in machine learning aims to update model weights and reduce training loss. Convergence represents the process of reaching a global optimum for gradient descent. We trained vessel segmentation models on RETA, DRIVE and UoA-DR datasets with the same baseline model VGAN and experimental settings. Each training set consists of 1600 images after augmenting by image rotation and flipping. Images in DRIVE and UoA-DR datasets are resized into the same image dimension as RETA. Optimizer is Adam with an initial learning rate of 2e-4 and loss function is binary crossentropy. Figure \ref{fig:5} plots mean and standard deviation of training loss over every 10 iterations in the first training epoch. RETA has a lower training loss across all iterations indicating less noisy pixel labels in its training set. Comparing training losses of the first and last iterations, UoA-DR shows the smallest speed of convergence than other datasets. This finding is consistent with subjective quality results in Table \ref{tab:4}. This approach could be an alternative indicator for vessel annotation quality.

\section*{Usage Notes}
RETA benchmark is licensed under a Creative Commons Attribution 4.0 International License. A proper citation of this work is expected if you use it in your work.

For vessel segmentation model training, we discourage users to resize provided 1024$\times$1024 binary vessel masks into other dimensions. Aliasing effect would generate disconnected small vessels and unsmooth vessel boundaries in a resized vessel mask even though antialiasing technique might alleviate it at some extent. RETA benchmark contains resized colour fundus images with the same size of binary vessel masks. User could also download original images from IDRiD Challenge and crop them into desired shape with our image transformation code. 

For evaluating vessel segmentation performance, accuracy, sensitivity, specificity and other pixel-wise matching based metrics are commonly used. However, these metrics show a limited fairness between computer predicted and manual labelled vessel images because of well known annotator variability problem. An instance is varied vessel masks (both in vessel thickness and location) labeled by the first and the second observers in DRIVE dataset. Thus, segment-level metrics like skeletal similarity\cite{yan2017skeletal} and topological similarity index \cite{araujo2021topological} that based on vessel skeleton and insensitive to vessel thickness might be more fair metrics on vessel segmentation evaluation. 

\begin{figure}
     \centering
     \begin{subfigure}[b]{0.32\textwidth}
         \centering
         \includegraphics[width=\textwidth]{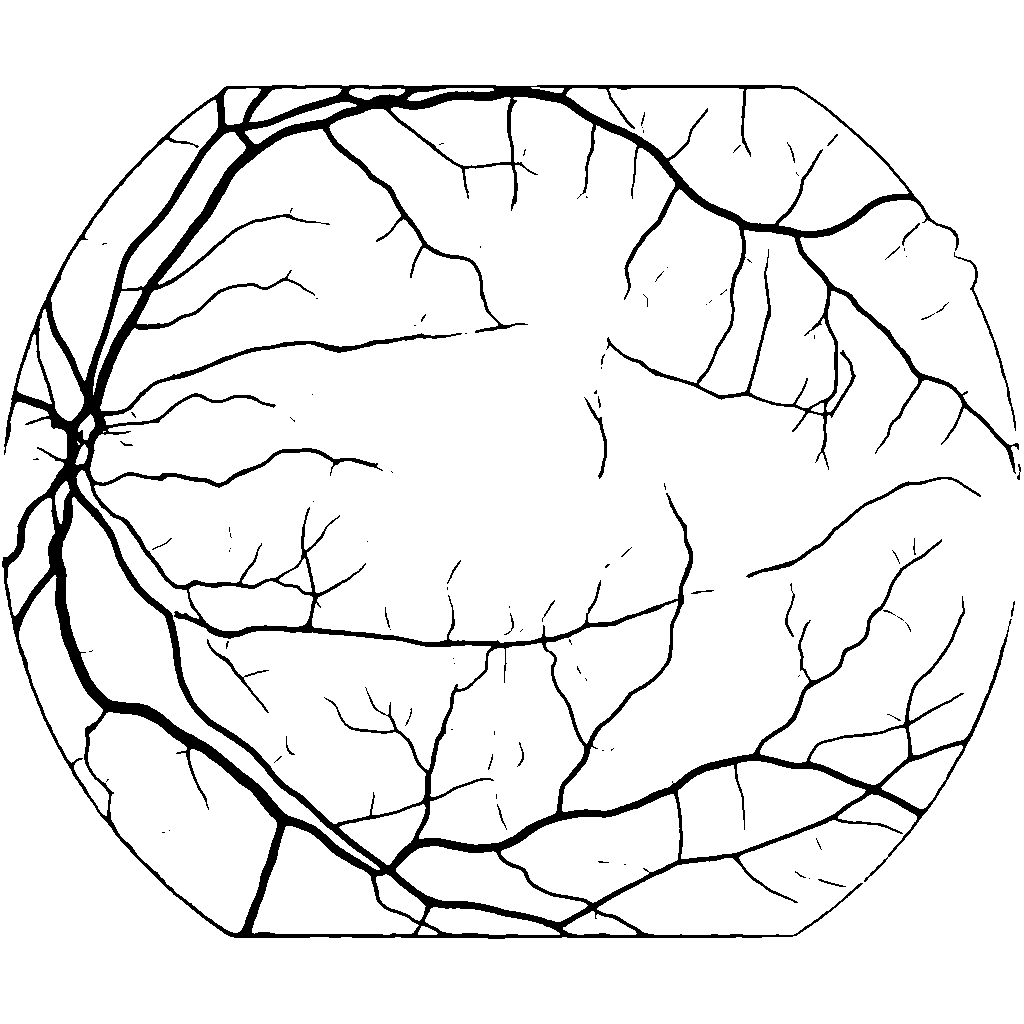}
         \caption{}
     \end{subfigure}
     \hfill
     \begin{subfigure}[b]{0.32\textwidth}
         \centering
         \includegraphics[width=\textwidth]{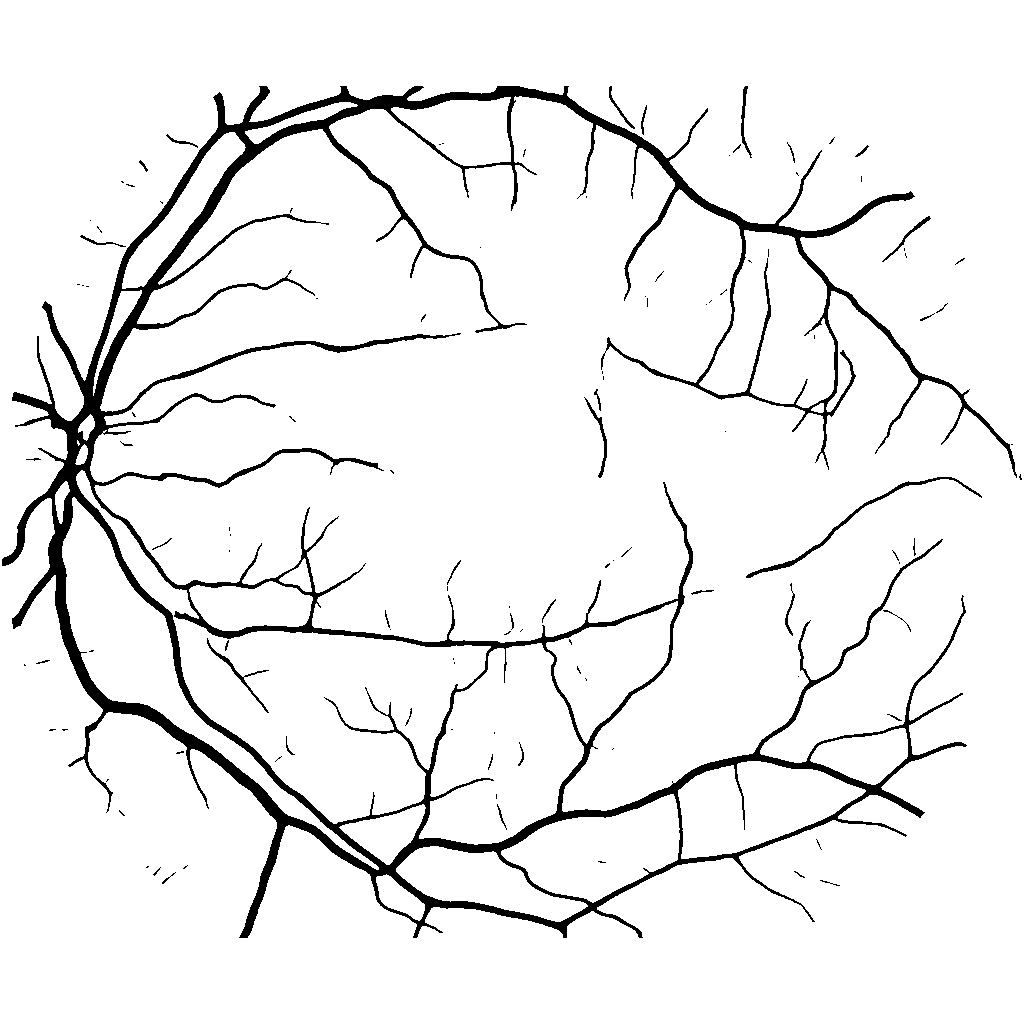}
         \caption{}
     \end{subfigure}
     \hfill
     \begin{subfigure}[b]{0.32\textwidth}
         \centering
         \includegraphics[width=\textwidth]{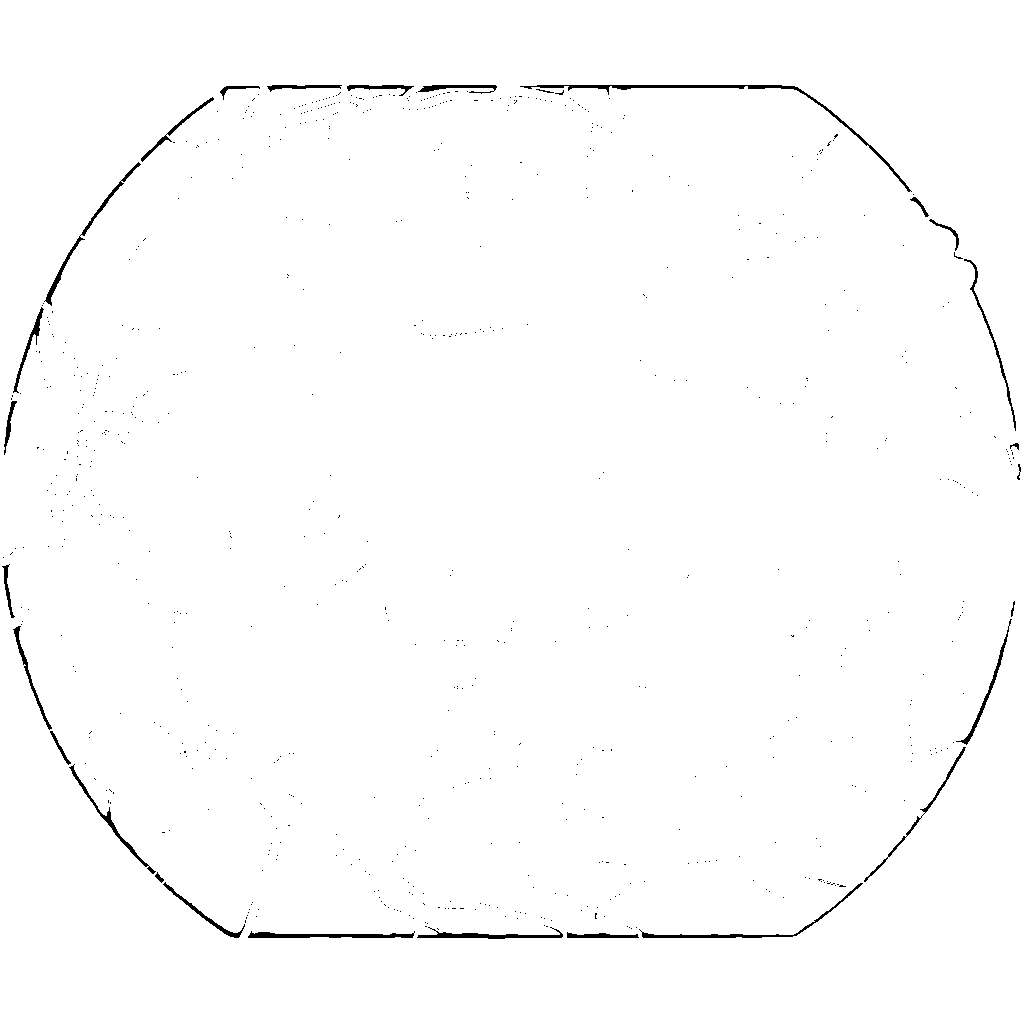}
         \caption{}
     \end{subfigure}
    \caption{Effectiveness of image preprocessing technique to remove border artefacts. Binary vessel segmentation images for the first (a)  original color image and (b) preprocessed color image of RETA testing set, (c) difference image between (a) and (b).}
    \label{fig:6}
\end{figure}

Another issue is false vessels detected around FOV border in model inference stage (one example is in Figure \ref{fig:6}(a)). It might be caused by strong contrast since pixel intensity outside FOV is zero. We followed Soares's preprocessing algorithm\cite{soares2006retinal} and padded artificial pixels around FOV. A MATLAB implementation named as 'FOV\_border\_extension.m' is available in the "codes" folder. Figure \ref{fig:6}(b) shows vessel segmentation result of preprocessed image. Compared with Figure \ref{fig:6}(a), the border artefacts are effectively removed. 

\section*{Code availability}
We released the following software, algorithms and models that used in the generation process of RETA benchmark. They are respectively saved in “codes”, “software” and “models” folders. Users could access them from our website.

\begin{itemize}
    \item CARL software. It could help users to visualize vessel annotations in a zoomed view. Users are also encouraged to evaluate segmentation performance of retinal anatomical structures and lesions within CARL.
    \item Image transformation and enhancement algorithm. The "codes" folder contains essential MATLAB source codes. Users can use "IDRiD\_cropImage.m" and "IDRiD\_restoreImage.m" to crop full-sized images of IDRiD challenge or restore a cropped vessel mask into original size. Script file "CARL\_image2mat.m" can transform users’ private images into custom mat files as inputs for CARL software.
    \item Vessel segmentation. A pretrained model and inference code are available at “models” folder. Please refer to usage document before running this code.
    \item Vessel centreline extraction. We utilised a python implementation of skeletonization method from scikit-image package to extract vessel centreline. The method parameter for skeletonize function is set as “lee”. 
\end{itemize}

\bibliography{main}

\section*{Acknowledgements}
This work was partly supported by the National Key R\&D Program of China (No. 2017YFB1002605). We appreciated Xin Ji, Zhijun Wang, Huanhui Wu, Yinling Du and Biling Weng from Shanggong Medical Technology $Co. Ltd$. for their assistance of this work. Ji and Wang provided us with additional financial aid and computational resource. Wu evaluated vessel annotation quality for 12 datasets. Du and Weng graded severity level of DR and DME of 81 fuduns images. They also proposed useful suggestions for developing CARL software. We would like to acknowledge Doctor Yi Zheng from Beijing Tongren Hospital, Capital Medical University for his work in final validation of vessel annotations. Finally, we acknowledged Prasanna Porwal and his team from India shared valuable IDRiD dataset.

\section*{Author contributions statement}
X.L. drafted the manuscript. He also developed CARL software and contributed to manual annotations of RETA benchmark. L.C. assisted to design this work and revised manuscript. S.Z. performed on project management and manuscript proofreading.

\section*{Competing interests} 
The authors declare no competing interests. 

\end{document}